\documentclass[usegraphicx]{mn2e}
\usepackage{psfig}

\catcode`\@=11
\def\gsim{\ifmmode{\mathrel{\mathpalette\@versim>}}
    \else{$\mathrel{\mathpalette\@versim>}$}\fi}
\def\lsim{\ifmmode{\mathrel{\mathpalette\@versim<}}
    \else{$\mathrel{\mathpalette\@versim<}$}\fi}
\def\@versim#1#2{\lower 2.9truept \vbox{\baselineskip 0pt \lineskip
    0.5truept \ialign{$\m@th#1\hfil##\hfil$\crcr#2\crcr\sim\crcr}}}
\catcode`\@=12
\renewcommand\[{\begin{equation}}
\renewcommand\]{\end{equation}}
\newcommand{\az}{a_0}
\newcommand{\gv}{{\bf g}}
\newcommand{\gvN}{\gv_{\rm N}}
\newcommand{\xv}{{\bf x}}
\newcommand{\phiN}{\phi_{\rm N}}
\newcommand{\hl}{h_{\lambda}}
\newcommand{\hm}{h_{\mu}}
\newcommand{\hn}{h_{\nu}}
\newcommand{\evl}{{\bf e}_{\lambda}}
\newcommand{\evm}{{\bf e}_{\mu}}
\newcommand{\evn}{{\bf e}_{\nu}}

\newcommand{\partiall}{\partial\lambda}
\newcommand{\partialm}{\partial\mu}
\newcommand{\partialn}{\partial\nu}

\newcommand{\nablasl}{\nabla^2_{\lambda}}
\newcommand{\nablasm}{\nabla^2_{\mu}}
\newcommand{\nablasn}{\nabla^2_{\nu}}

\newcommand{\Dphi}{{\cal D}_{\phi}}
\newcommand{\Ao}{{\cal A}_1}
\newcommand{\Bo}{{\cal B}_1}
\newcommand{\Co}{{\cal C}_1}

\newcommand{\Ad}{{\cal A}_2}
\newcommand{\Bd}{{\cal B}_2}
\newcommand{\Cd}{{\cal C}_2}

\newcommand{\At}{{\cal A}_3}
\newcommand{\Bt}{{\cal B}_3}
\newcommand{\Ct}{{\cal C}_3}

\newcommand{\Nor}{{\cal N}}

\newcommand{\gN}{g_{\rm N}}
\newcommand{\hv}{{\bf h}}

\newcommand{\Sv}{{\bf S}}

\newcommand{\be}{\begin{equation}}
\newcommand{\ee}{\end{equation}}

   \title[Separable MOND potential-density pairs]
         {Separable triaxial potential-density pairs in MOND}
        
   \author[Ciotti, Zhao \& de Zeeuw]
          {Luca Ciotti$^1$, Hongsheng Zhao$^{2,3}$, P. Tim de Zeeuw$^{4,5}$
           \\ $^1$Astronomy Department, University of Bologna, 
              via Ranzani 1, 40127 Bologna, Italy
           \\ $^2$ SUPA, University of St Andrews, KY16 9SS, UK
           \\ $^3$ Vrije Universiteit, De Boelelaan 1081, 1081 HV Amsterdam, The Netherlands
           \\ $^4$Sterrewacht Leiden, Universiteit Leiden, Postbus 9513, 2300 RA Leiden, 
                 The Netherlands
           \\ $^5$European Southern Observatory, Karl-Schwarzschild-Str 2, 85748 Garching, Germany
           }

\date{Accepted 2012 February 7. Received 2012 February 5; in original
  form 2011 September 27}
\pubyear{2012}
 
\begin{document} 
\maketitle

\begin{abstract} 

  We study mass models that correspond to MOND (triaxial) potentials
  for which the Hamilton-Jacobi equation separates in ellipsoidal
  coordinates.  The problem is first discussed in the simpler case of
  deep-MOND systems, and then generalized to the full MOND regime.  We
  prove that the Kuzmin property for Newtonian gravity still holds,
  i.e., that the density distribution of separable potentials is fully
  determined once the density profile along the minor axis is
  assigned. At variance with the Newtonian case, the fact that a
  positive density along the minor axis leads to a positive density
  everywhere remains unproven. We also prove that (i) all regular
  separable models in MOND have a vanishing density at the origin, so
  that they would correspond to centrally dark-matter dominated
  systems in Newtonian gravity; (ii) triaxial separable potentials
  regular at large radii and associated with finite total mass leads
  to density distributions that at large radii are not spherical and
  decline as $\ln (r) /r^5$; (iii) when the triaxial potentials admit
  a genuine Frobenius expansion with exponent $0<\epsilon <1$, the
  density distributions become spherical at large radii, with the
  profile $\ln(r)/r^{3+2\epsilon}$.  After presenting a suite of
  positive density distributions associated with MOND separable
  potentials, we also consider the important family of (non-separable)
  triaxial potentials $V_1$ introduced by de Zeeuw \& Pfenniger, and
  we show that, as already known for Newtonian gravity, they obey the
  Kuzmin property also in MOND. The ordinary differential equation
  relating their potential and density along the $z$-axis is an Abel
  equation of the second kind that, in the oblate case, can be
  explicitly reduced to canonical form.


\end{abstract}

\begin{keywords}
galaxies: kinematics and dynamics --- galaxies: structure --- dark
matter --- methods: analytical --- stellar dynamics

\end{keywords}

\section{Introduction}
\label{secint}

Milgrom (1983) proposed that the failure of galactic rotation curves
to decline in Keplerian fashion outside the galaxies' luminous body
arises not because galaxies are embedded in massive dark halos obeying
Newtonian gravity, but because Newton's law of gravity has to be
modified for fields that generate accelerations smaller than some
characteristic value $\az\simeq 1.2\times 10^{-8}$ cm s$^{-2}$.
Subsequently, in order to solve basic problems presented by this
phenomenological formulation of the theory (now known as Modified
Newtonian Dynamics or MOND), such as conservation of linear momentum
(e.g., Felten 1984), Bekenstein \& Milgrom (1984) substituted the heuristic
model with the MOND non-relativistic field equation
\[
\nabla\cdot\left[\mu\left({\Vert\nabla\phi\Vert\over\az}\right)
                 \nabla\phi\right] = 4\pi G \rho,
\label{eqMOND}
\]
where $\Vert ...\Vert$ is the standard Euclidean norm, $\mu$ is a
scalar function, $\phi$ is the gravitational potential produced by the
density distribution $\rho$, and
\[
\gv=-\nabla\phi,
\label{eqgv}
\]
is the MOND gravitational field experienced by a test particle.
For an isolated system of finite mass, eq.~(\ref{eqMOND}) is
supplemented with the natural boundary condition $\nabla\phi\to 0$ for
$\Vert\xv\Vert\to\infty$.  Equation~(\ref{eqMOND}) is obtained from a
variational principle applied to a Lagrangian with all the required
symmetries, so the standard conservation laws are obeyed\footnote{An
  alternative non-relativistic formulation of MOND, dubbed QMOND, has
  been proposed (Milgrom 2010), but it is not discussed here.}.  In
the regime of intermediate accelerations the function $\mu$ is not
fully constrained by theory or observations, while asymptotically
\[
\mu(t)\sim\cases{t&for $t\ll 1$,\cr 1&for $t\gg 1$. }
\label{eqmuasym}
\]
A common choice is
\[
\mu (t)={t\over\sqrt{1+t^2}},
\label{eqmu}
\]
(see also Famaey \& Binney 2005, Zhao \& Famaey 2006).

From eq.~(\ref{eqmuasym}) it follows that eq.~(\ref{eqMOND})
reduces to the Poisson equation when $\Vert \nabla \phi \Vert \gg
\az$, while the limit equation 
\[
\nabla\cdot\left({\Vert\nabla\phi\Vert}\nabla\phi\right) = 4\pi G \az
\rho,
\label{eqdMOND}
\]
describes systems for which (or regions of space where)
$\Vert\nabla\phi\Vert \ll\az$, i.e. systems for which the MOND
predictions differ most from the Newtonian
ones. Equation~(\ref{eqdMOND}), characterizing the so-called ``deep
MOND regime'' (hereafter dMOND), is then of particular relevance in
MOND investigations, as this is the regime where one hopes to find the
most significant differences with the predictions of Newtonian
gravity.  From the mathematical point of view, the l.h.s. of
eq.~(\ref{eqdMOND}) is a special case of the so-called $p$-Laplace
operator $\nabla(\Vert\nabla\phi\Vert^{p-2}\nabla\phi)$.  The dMOND
case corresponds to $p=3$, the so-called critical case for $\Re^3$.  A
large body of mathematical literature is dedicated to the
$p$-Laplacian, but due to its non-linearity it is not surprising that
several important questions are still open (e.g., see Lindqvist \&
Manfredi 2008). We anticipate that the general results in the present
paper are independent of the specific form of $\mu$, because they just
follow from the fact that $\mu$ is a function of
$\Vert\nabla\phi\Vert$, or they are obtained for the dMOND regime.

A considerable body of observational data seems to support MOND well
beyond its originally intended field of application (see, e.g.,
Milgrom 2002, Sanders \& McGaugh 2002, Famaey \& McGaugh 2011), but
potential problems of the theory have been pointed out by many authors
(see, e.g. The \& White 1988, Buote et al.~2002, Sanders 2003, Ciotti
\& Binney 2004, Knebe \& Gibson 2004, Zhao et al.~2005, Ibata et
al.~2011, Galianni et al.~2011). At present, the situation is in general considered
unsettled.  It is thus natural to study in detail MOND predictions, in
particular focusing on dMOND systems, i.e. systems that in Newtonian
gravity would be dark matter dominated.  Unfortunately, MOND
investigations, especially on the theory side, have been considerably
slowed down by the almost complete lack of aspherical
density-potential pairs, needed to test the predictions in cases more
realistic than those described by spherical symmetry.  In MOND, the
main difficulty to obtain exact aspherical density-potential pairs
originates from the fact that a simple relation between the Newtonian
and the MOND gravity fields, produced by an assigned density
distribution, in general does not exist.  In fact, the Newtonian
potential $\phiN$ obeys the Poisson equation
\[
\nabla^2\phiN=4\pi G\rho, 
\label{eqPoisson}
\]
so that $\mu(\Vert\nabla\phi\Vert/\az)\nabla\phi$ and $\nabla\phiN$
differ, for assigned $\rho$, by a solenoidal field $\Sv={\rm
  curl\,}\hv$. In turn, the potential vector $\hv$ depends on $\rho$,
and so it is apriori unknown.  The only exception is provided by
density distributions in which the modulus $\gN=\Vert\gvN\Vert$ of the
Newtonian field $\gvN =-\nabla\phiN$ is stratified on surfaces of
constant $\phiN$ (particularly simple cases are those of spherically
and cylindrically symmetric densities, or densities stratified on
homogeneous planes, see also Brada \& Milgrom 1995; Shan et
al.~2008). In such cases $\Sv$ vanishes, and
\[
\mu\left({\Vert \gv\Vert \over\az}\right)\gv=\gvN .
\label{eqmug}
\]
Equation~(\ref{eqmug}) coincides with the original MOND formulation of
Milgrom (1983) and can be solved algebraically for $\gv$ in terms of
$\gvN$, just by taking its norm.

A general method to build aspherical and exact MOND density potential
pairs is presented in Ciotti et al. (2006) where, by
extending the homeoidal expansion technique introduced in Ciotti \&
Bertin (2005) for Newtonian gravity, it is shown how a ``seed''
spherical potential can be deformed to an axisymmetric or triaxial
shape, leading to analytical density-potential pairs satisfying the
MOND equation. Unfortunately, the resulting density distributions are
not fully under control in case of major departures from spherical
symmetry, and regions of unphysical negative density may result.

For these reasons here we explore a different approach, i.e. we focus
on the possibility to extend to MOND some of the remarkable results
obtained in Newtonian gravity for potentials separable in ellipsoidal
coordinates. This not only to better understand the properties of MOND
systems, but also to develop a new method to generate exact solutions
deviating from spherical symmetry for the $p$-Laplace operator.  While
we refer to other papers for the full account of separability in
ellipsoidal coordinates (de Zeeuw 1985b, hereafter Z85; de Zeeuw \&
Lynden-Bell 1985, hereafter ZLB85), here we just summarize the
properties of Newtonian separable potentials in ellipsoidal
coordinates relevant to the present investigation. Some of these
properties are indeed shared by similar but non-separable families
described in de Zeeuw \& Pfenniger (1988, hereafter ZP88).
Specifically, in Newtonian separable systems:

1. The potential along the long axis of the coordinate ellipsoids (the
$z$-axis in the standard convention) is related to the density profile
along this axis by a linear second order ordinary differential
equation (ODE). This is the {\it Kuzmin property}.

2.  For assigned density profile along the $z$-axis, the ODE can be
integrated completely.  Once the parameters of the ellipsoidal
coordinate systems are fixed, the solution determines uniquely the
potential (and so the density) over the whole space. The density
elsewhere is related to that on the $z$-axis by the so-called {\it
  Kuzmin formula}.  Usually, the $z$-axis is the {\it short} axis of
the density distribution.

3. The Kuzmin formula shows that the density is everywhere
non-negative if this holds along the short axis.  This is the {\it
  Kuzmin theorem}.

4.  The Kuzmin formula also shows that density profiles that fall off
along the $z$-axis faster than $z^{-3}$ lead to finite mass. Density
profiles that fall off less steep than $z^{-4}$ become spherical at
large radii, while for $\rho(z)\propto z^{-4}$ or steeper the models
have finite flattening at large radii. In particular, density profiles
that fall off faster than $z^{-4}$ lead to density distributions that
in all other directions falls off as $r^{-4}$, so that such models are
quasi-toroidal (de Zeeuw, Peletier \& Franx 1986; ZP88).

Of course, separable potentials are a special - albeit very important
- class of potentials expressed in ellipsoidal coordinates. The
interest in separable potentials is that, once a supporting positive
density can be found, then their orbital classification can be done
exactly, and equally well in MOND or in Newtonian gravity. In
addition, as we will briefly discuss in the Conclusions, separable
potentials may allow for a contructive approach towards the assembly
of self-consistent MOND modes, i.e., collisionless systems supported
by a positive phase-space distribution function obeying the Jeans
Theorem (e.g., Lynden-Bell 1962, Binney \& Tremaine 2008).  However,
when considering the more general problem of obtaining a flexible
approach to the construction of triaxial MOND potential-density pairs,
different classes of (non-separable) potentials still obeying the
Kuzmin property are worth to be explored, such as those described in
ZP88. These models are not separable, but their simpler algebrical
structure leads to simpler equations, that might be solved
(numerically) more easily than in the separable case.

The paper is organized as follows. In Section 2 we derive the ODE
along the $z$-axis for separable models in the dMOND regime, and then
we extend the result to the full MOND case, thus showing that the
Kuzmin property holds in MOND. In Section 3 we derive the general
asymptotic behavior at the center and at large radii of the density
distributions generated by MOND regular separable potentials. Some
explicit examples of everywhere positive densities associated with
separable potentials are then presented in Section 4.  A discussion of
a special family of non-separable triaxial potentials (the $V_1$
family of ZP88) is carried out in Section 5, where among other
findings it is shown that the Kuzmin property holds also for $V_1$
systems.  The main conclusions are summarized in Section 6.  In the
Appendix we list technical details, together with a brief discussion
of the separable axisymmetric power-law models by Sridhar \& Touma
(1997) in the context of MOND.

\section{The Kuzmin property for separable MOND systems}

We consider mass models that correspond to
(triaxial) MOND potentials for which the Hamilton-Jacobi equation
separates in ellipsoidal coordinates $(\lambda, \mu, \nu)$, and 
investigate whether a Kuzmin formula holds for them.  This is not
obvious, as the MOND field equation is non-linear and considerably
more complicated than the Poisson equation.  We recall that the most
general form of a separable potential in ellipsoidal coordinates 
can be written as
\[
\phi= -{F(\lambda)\over (\lambda -\mu)(\lambda -\nu)}
      -{F(\mu)\over (\mu -\nu)(\mu -\lambda)}
      -{F(\nu)\over (\nu -\lambda)(\nu -\mu)},
\label{phisep}
\]
where $F(\tau)$ is an arbitrary\footnote{In principle, three different
  functions $F_1(\lambda)$, $F_2(\mu)$,  and $F_3(\nu)$ may be allowed
  in    separable    potentials.    Smooth   mass    models    require
  $F_1(-\alpha)=F_2(-\alpha)$  and  $F_2(-\beta)=F_3(-\beta)$ together
  with conditions on  the derivatives of these functions,  so it is in
  most  cases no  loss of  generality to  take  $F_1=F_2=F_3=F$ (e.g.,
  Lynden-Bell  1962,  ZLB85).}  function  (Z85, ZLB85);  the  relevant
properties  of   ellipsoidal  coordinates  needed   in  the  following
discussion are summarized in Appendix A.

It turns out that we can obtain information on the full MOND case just
by restricting to the simpler case of dMOND systems, i.e., by focusing
on the properties of the $p$-Laplacian.  First, we rewrite
eq.~(\ref{eqdMOND}) in the more convenient form
\[
4\pi G\az\rho = 
\Vert\nabla\phi\Vert\nabla^2\phi+{\Dphi\Vert\nabla\phi\Vert^2\over 
2\Vert\nabla\phi\Vert},
\label{dMONDell}
\]
where the explicit expression for the linear differential operator
$\Dphi\equiv \langle\nabla\phi,\nabla\rangle$ in terms of ellipsoidal
coordinates is given in eq.~(\ref{eq:dphi}).  The advantage of working
with eq.~(\ref{dMONDell}) instead of eq.~(\ref{eqdMOND}) is that one
avoids the computation of the derivatives of a norm, with the involved
square root.

In principle, by inserting eq.~(\ref{phisep}) in eq.~(\ref{dMONDell}),
and using the formulae reported in the Appendix, some heavy algebra
will give the expression for the density distribution over the whole
space as a function of $F$, and its first and second order derivative,
$F'$ and $F''$. Not unexpectedly, the resulting formula is quite
formidable, and of little use. In practice, given $F(\tau)$, the
explicit computation can be performed by using one of the many
available computer algebra packages, and some cases will be discussed
in Section 4.

Here instead we are interested in a more general property, i.e. if the
Kuzmin formula (or even the Kuzmin theorem) holds also for MOND
separable systems.  From eq.~(\ref{eq:ellcoo}) it follows that 
the density profile along the whole $z$-axis of a generic density
distribution $\rho(\lambda,\mu,\nu)$ is given by the function
\[
\psi(\tau)\equiv\cases{\rho(-\alpha,-\beta,\tau),
\quad -\gamma\leq\tau\leq -\beta\cr
\rho(-\alpha,\tau,-\beta),\quad -\beta\leq\tau\leq -\alpha\cr
\rho(\tau,-\alpha,-\beta),\quad -\alpha\leq\tau}
\label{eq:zaxis}
\]
where in the three intervals $z^2=\tau +\gamma$.  In analogy with Z85
(see also Kuzmin 1956 for the oblate axisymmetric case), we now show
that the r.h.s. of eq.~(\ref{dMONDell}), evaluated on the $z$-axis,
reduces in the three intervals of $\tau$ to the {\it same}
second-order ODE for the unknown function $F(\tau)$, thus proving that
also in dMOND the function $\psi(\tau)$ determines $F(\tau)$, and so
the density field everywhere.  In other words, {\it the Kuzmin
  property (and so a Kuzmin-like formula) holds also for separable
  dMOND systems}.

Let us consider the restriction of eq.~(\ref{dMONDell}) to
the $z$-axis.  The Laplace operator satisfies the Kuzmin
property, so there is nothing to prove, and its expression along the
$z$-axis is given in eq.~(27) of Z85.  An explicit computation then
shows that the restriction to the $z$-axis of $||\nabla\phi ||^2$ is
given by the unique expression
\[
||\nabla\phi ||_z^2=4(\tau+\gamma)\Nor^2,
\label{eq:snormz}
\]
in the three intervals spanned by $\tau$, where
\begin{eqnarray}
\Nor&=&
{F'(\tau)\over (\tau+\alpha)(\tau+\beta)}-
{(2\tau+\alpha+\beta)F(\tau)\over
(\tau+\alpha)^2(\tau+\beta)^2}+\cr
&&
{F(-\alpha)\over
(\tau+\alpha)^2(\beta-\alpha)}-
{F(-\beta)\over
(\tau+\beta)^2(\beta-\alpha)}.
\label{eq:normN}
\end{eqnarray}
Remarkably, we note that, with the exception of the factor
$2\sqrt{\tau+\gamma}$, no irrationalities are involved in the
expression of $||\nabla\phi ||_z$. Of course, care is needed in the
evaluation of this latter quantity, as it contains the absolute value
$|\Nor |$. Finally, some algebra shows that the restriction of
$\Dphi||\nabla\phi ||^2$ to the $z$-axis also admits the unique
representation
\[
[\Dphi||\nabla\phi ||^2]_z=-4||\nabla\phi ||_z^2
\left[ \Nor+2(\tau+\gamma){\cal M}\right ],
\label{eq:dnormz}
\]
where 
\begin{eqnarray}
{\cal M}&=&{d\Nor\over d\tau}=
{F''(\tau)\over (\tau+\alpha)(\tau+\beta)}-
2F'(\tau){2\tau+\alpha+\beta\over
(\tau+\alpha)^2(\tau+\beta)^2}+\cr
&&
2F(\tau){3\tau^2+3\tau (\alpha+\beta)+\alpha^2+\beta^2+\alpha\beta\over
(\tau+\alpha)^3(\tau+\beta)^3}-\cr
&&
{2F(-\alpha)\over
(\tau+\alpha)^3(\beta-\alpha)}+
{2F(-\beta)\over
(\tau+\beta)^3(\beta-\alpha)}.
\label{eq:normM}
\end{eqnarray}
Combining the previous results, we obtain the following second-order
ODE relating, for separable dMOND systems, the density profile along
the whole $z$-axis to $F(\tau)$:
\[
2\pi G\az\psi(\tau)=\sqrt{\tau+\gamma}\;|{\cal N}|
                                \left\{
                               [\nabla^2\phi]_z-2{\cal N}-4(\tau+\gamma){\cal M}
                                \right\}.
\label{eq:ODEz}
\]
Thus, the preliminary result is that the Kuzmin property holds in
dMOND (i.e., for the $p$-Laplacian).  {\it Furthermore, with the aid
  of the previous results it follows immediately, by restriction of
  eq.~(\ref{eqMOND}) to the $z$-axis, that the Kuzmin property (and in
  principle a Kuzmin formula) also holds for the full MOND field
  equation, as the $\mu$ function in eq.~(\ref{eqMOND}) depends on
  $\Vert\nabla\phi\Vert$}.  Unfortunately, the non-linearity of the
problem seems to prevent the construction of the explicit Kuzmin
formula even in dMOND, so that the successive analysis of positivity
of the density as in Newtonian gravity (Z85) cannot be performed, and
the Kuzmin theorem remains unproven.

We conclude this general Section by noticing that, as pointed out by
the Referee, a more general result on Kuzmin property can in fact be
obtained, encopassing the present results and those in Section 5. In
practice, with some analytical work, it can be shown that the Kuzmin
property certainly hold in Newtonian gravity and in MOND for any
potential that can be written as a symmetric function of ellipsoidal
coordinates, i.e., by a function invariant for the transformation
$\lambda\to\mu\to\nu$.

\section{Asymptotic behaviors}

In the previous Section we proved that MOND systems with separable
potentials in ellipsoidal coordinates obey the Kuzmin property, and so
in principle a Kuzmin formula holds for them.  Before embarking on the
study of the density distributions associated with specific separable
potentials, we focus on the more general question of the asymptotic
behavior of the density, both at the center and at large radii (for
systems with finite total mass).

\subsection{Behavior at the center}

For a (regular) function $F(\tau)$ we begin by considering its
second-order Taylor expansion near the center, i.e.
\begin{eqnarray}
\cases{
\displaystyle{F(\lambda)\sim 
F(-\alpha)+F'(-\alpha)(\lambda+\alpha)+
{F''(-\alpha)(\lambda+\alpha)^2\over 2}},\cr
\displaystyle{F(\mu)\sim
F(-\beta)+F'(-\beta)(\mu+\beta)+
{F''(-\beta)(\mu+\beta)^2\over 2}},\cr
\displaystyle{F(\nu)\sim
F(-\gamma)+F'(-\gamma)(\nu+\gamma)+
{F''(-\gamma)(\nu+\gamma)^2\over2}}.
}
\label{eq:phiexp}
\end{eqnarray}
In the expansion above we limit to second order, but the present
analysis can be carried out (with increasing algebraically complexity)
to any desired order\footnote{For example, all the results in this
  Section have been re-obtained by using a Taylor series for $F(\tau)$
  truncated at the $10^{th}$ order (inclusive), and performing the
  expansions with Mathematica.}.  Before embarking on the following
discussion, it is important to recall that the function $F$ allows for
a linear gauge i.e., two functions differing for $a\tau+b$, with $a$
and $b$ constants, lead to the same function $\phi$ in
eq.~(\ref{phisep}), as can be easily verified by direct substitution.
With the aid of this gauge it is possible to assign two prescribed
values to $F$ at two arbitrary points, for example to impose that
$F(-\alpha)=F(-\gamma)=0$, so that for regular $F$ one can write
$F(\tau)=(\tau+\alpha)(\tau+\gamma)G(\tau)$. This form has been proved
very useful in several investigations (e.g., de Zeeuw 1985a, Z85,
Hunter \& de Zeeuw 1992; Arnold, de Zeeuw \& Hunter 1994; van de Ven
et al. 2003). Here we refrain from using the factorized form, and all
the formulae are given in full generality: of course, more compact (but
less symmetric) expressions in terms of the function $G$ can be
immediately found from those reported here, by simple algebraical
substitution.

We focus first on the dMOND regime, recalling that near the center
$\lambda+\alpha\sim x^2$, and analogous relations hold for the $\mu$
and $\nu$ coordinates (see eq.~[\ref{eq:centcor}]). We consider the
behavior of the different operators at the r.h.s of
eq.~(\ref{dMONDell}), beginning with the Laplacian.  As is well known,
the application of the Laplace operator to a regular separable
potential leads to a finite central value
\[
{\nabla^2\phi_0\over 2} = {
(\gamma-\beta)F'(-\alpha)+
(\alpha-\gamma)F'(-\beta)+
(\beta-\alpha)F'(-\gamma)\over
\triangle},
\label{eq:lapcen1}
\]
where
\[
\triangle\equiv (\alpha-\beta)(\alpha-\gamma)(\beta-\gamma)<0.
\label{eq:tria}
\]
(Z85, eq.~[27]).  Therefore, barring the special case of null
derivatives of $F$ at $\tau=-\alpha$ ,$-\beta$, and $-\gamma$ (or the
more general case discussed later), according to eq.~(\ref{eqPoisson})
the value of the central density is non-zero in separable Newtonian
systems.

We now move to the term $||\nabla\phi||$, noticing that it appears in
the denominator of eq.~(\ref{dMONDell}), and so the convergence of the
density near the origin may be not guaranteed in case of a vanishing
gradient left unbalanced by the behavior of the term $\Dphi
||\nabla\phi||^2$. Indeed, the vanishing of $||\nabla\phi||$
at the center of a regular triaxial potential is expected from
geometrical considerations, and in fact, by using
eq.~(\ref{eq:phiexp}) we find that near the origin
\[
||\nabla\phi||^2\sim 4(\Ao^2x^2+\Bo^2y^2+\Co^2z^2),
\label{eq:normcen1}
\]
where the three constants $\Ao$, $\Bo$, and $\Co$ depend on $\alpha$,
$\beta$ and $\gamma$, and their explicit expression is given in
eqs.~(\ref{eq:coecen1a})-(\ref{eq:coecen1b}).  Note that the
expression above is positive, because the expanded
function is positive definite, and the higher order terms cannot
affect the sign for sufficiently small displacements from the origin.
In particular, as the three ellipsoidal coordinates are independent,
the three coefficients must be positive, and in fact they are perfect
squares. More generally, a similar positivity argument holds for the
leading term in the expansion of $||\nabla\phi||^2$ independently of
the order, i.e., the first {\it non-zero} term in the expansion is
necessarily positive near the origin, as we will show in the following.

We now focus on the last term in eq.~(\ref{dMONDell}). After some
computation, it is found that near the origin
\[
\Dphi ||\nabla\phi||^2\sim 16 (\Ao^3x^2+\Bo^3y^2+\Co^3 z^2),
\label{eq:dnormcen1}
\]
where the three coefficients are the same as in
eq.~(\ref{eq:normcen1}).

Finally, by combining the previous results, a simple calculation shows
that in general near the origin
\[
\rho\sim {\Ao^2(2\Ao+\nabla^2\phi_0)x^2 +
  \Bo^2(2\Bo+\nabla^2\phi_0)y^2 + \Co^2(2\Co+\nabla^2\phi_0)z^2\over 
2\pi G\az\sqrt{\Ao^2 x^2 + \Bo^2 y^2 + \Co^2 z^2}},
\]
A change to spherical coordinates then proves that {\it the central
  density of dMOND separable systems vanishes, linearly with the
  spherical radius $r$. }

A natural question arises, i.e. can we say something about $\rho$ in
the special circumstance of $\Ao =\Bo =\Co =0$? It could be that the
order balance between the numerator and the denominator in
eq.~(\ref{dMONDell}) breaks down when the leading term of
$||\nabla\phi||^2$ near the origin is of higher order.  The obtained
results are indeed quite interesting. First of all, by inspection of
eqs.~(\ref{eq:coecen1a})-(\ref{eq:coecen1b}), it follows that the
condition $\Ao=\Bo=\Co=0$ is equivalent to the requirement that
$F'(-\alpha)$, $F'(-\beta)$, and $F'(-\gamma)$ are well defined
functions of $\alpha$, $\beta$, $\gamma$, and of $F(-\alpha)$,
$F(-\beta)$, $F(-\gamma)$.  Accordingly, we consider the potential in
eq.~ (\ref{eq:phiexp}), with the values of $F'$ fixed by the special
case just described (and increasing the adopted order of expansion of
$F$ to the third one).  The leading terms of the expansions now read
\begin{eqnarray}
\cases{
\nabla^2\phi\sim 6 (\Ad x^2+\Bd y^2+\Cd z^2),\cr
||\nabla\phi||^2\sim 4(\Ad^2x^6+\Bd^2y^6+\Cd^2z^6),\cr
\Dphi ||\nabla\phi||^2\sim 48(\Ad^3 x^8+\Bd^3 y^8+\Cd^3 z^8),
\label{eq:dencen2}
}
\end{eqnarray}
where the explicit form of the coefficients is given in
eqs.~(\ref{eq:coecen2a})-(\ref{eq:coecen2b}).  Note that in this case
also the Laplace operator vanishes at the orgin, and a simple
calculation shows that the density near the center vanishes as $r^5$.

We finally repeat the argument above, with the additional request that
also $\Ad=\Bd=\Cd=0$, thus fixing the values of $F''(-\alpha)$,
$F''(-\beta)$, and $F''(-\gamma)$ from
eqs.~(\ref{eq:coecen2a})-(\ref{eq:coecen2b}).  By using
eq.~(\ref{eq:phiexp}) with $F(\tau)$ expanded up to the fourth order
inclusive, we now find
\begin{eqnarray}
\cases{
\nabla^2\phi\sim 5 (\At x^4+\Bt y^4+\Ct z^4),\cr
||\nabla\phi||^2\sim \At^2x^{10}+\Bt^2y^{10}+\Ct^2z^{10},\cr
\Dphi ||\nabla\phi||^2\sim 10(\At^3 x^{14}+\Bt^3 y^{14}+\Ct^3 z^{14}),
\label{eq:dencen3}
}
\end{eqnarray}
where the coefficients are given in eqs.~(\ref{eq:coecen3}): now the
density at the center vanishes as $r^9$. By repeating this exploration
to higher and higher orders, we find that the order of vanishing of
the density, as a function of radius $r$, increases by 4 for each
additional order of regularity imposed on $\Vert\nabla\phi\Vert$ near
the origin.  We also found that, from the third order upward, the
first non-zero coefficients ${\cal A}_k$, ${\cal B}_k$, and ${\cal
  C}_k$ of the leading terms near the origin depend only on the
derivatives $F^{(k)}(\tau)$ evaluated at $-\alpha$, $-\beta$ and
$-\gamma$, so that high-order regularity at the center can be
expressed directly as the vanishing of the corresponding derivatives
of $F$ at the origin.

Therefore, the previous analysis shows that a generic dMOND system
associated with a regular triaxial separable potential has - at
variance with Newtonian gravity - a zero density at the center. This
fact leads to some non-trivial consequences. The first is that this
result remains true even when using the full MOND equation, as can be
verified with a formal expansion. A simple argument is as follows. If
we use the full MOND equation, and the central regions are not in
dMOND regime, then they are described by Newtonian gravity. But the
Newtonian force in triaxial regular separable potentials vanishes, so
{\it all} MOND separable systems at the center are actually in dMOND
regime, with the consequent vanishing central density.  Of course, the
vanishing of the central density in MOND may well occur also for other
families of non-separable potentials (e.g., for potentials with a
sufficient degree of reflection symmetries along the coordinate axes).

The second consequence follows from a further argument. Suppose MOND
holds, and consider a separable system of baryonic density $\rho$, so
that from the previous result $\rho =0$ at the origin. We now focus on
the the {\it total} density $\rho_{\rm N}$ of the so-called Equivalent
Newtonian System associated with the baryonic density $\rho$, i.e.,
the mass distribution needed in Newtonian gravity to produce the same
gravity field of MOND. Of course, $\rho_{\rm N}$ is obtained by
application of the Laplace operator to the MOND potential, so that in
the Newtonian framework the baryonic density $\rho$ results
``immersed'' in a dark matter halo of density $\rho_{\rm
  h}\equiv\rho_{\rm N}-\rho$, and from eq.~(17) the halo density at
the center will be different from zero.  Provided $\rho_{\rm h}\geq 0$
everywhere (a non trivial request), we are lead to conclude that {\it
  a separable MOND system would appear, when interpreted in the
  context of Newtonian gravity, fully dark matter dominated at the
  center}, with an arbitrarily large (formally infinite) mass-to-light
ratio near the origin.  Note that this property may be expected also
in other families of MOND potentials, not necessarily separable.  In
fact, it can be easily proved that a regular potential with reflection
symmetries - $\phi=\phi(x^2,y^2,z^2)$ - leads to a density with an
expansion near the center identical to eq.~(21), independently of
separability. However, at higher orders the special form of
eqs.~(22)-(23) is not obtained, showing that separability removes the
cross-terms and leaves diagonalized quartics, sextics, and so on.

We conclude by noting that the addition of a central black hole would
change the central force field from dMOND to Newtonian, in principle
opening the possibility to have systems with a non-zero central
density. Unfortunately, the addition of a central mass breaks down
separability of triaxial potentials (excluding exceptional
axisymmetric cases, such that discussed in Appendix B).

\subsection{Behavior at large radii}

The other place where the asymptotic analysis can be carried out in
generality is at infinity. In particular, from
eqs.~(\ref{eq:ellipse})-(\ref{eq:ellcoo}), it follows that
$\lambda\sim r^2$ for $r\to\infty$, with $r$ being the radius in
spherical coordinates.  In order to better illustrate the MOND case,
we begin by recalling the idea behind the computation in Newtonian
gravity.  For a system of finite total mass $M$, Newtonian gravity
dictates that $\phi\sim -GM/r$ for $r\to\infty$. If we ask also for
separability, then it is easy to show that the required asymptotic
trend is matched in eq.~(\ref{phisep}) if and only if
\[
F(\tau)=\tau^{3/2}h(\tau),\quad h(\tau)\sim 1\quad {\rm for}\; \tau\to\infty.
\label{Fna}
\]
Note that in the expression above we fixed $GM=1$: it is simple to
restore this dimensional factor in the obtained density distribution
after the application of the Laplace operator. From now on we refer to
$h(\tau)$ as to the {\it shape function}: its relevance in determining
the mass profile at large radii will be discussed in detail in the
dMOND context.

By evaluating the Laplace operator, one recovers the well known result
of Newtonian gravity (e.g. de Zeeuw, Franx \& Peletier 1986) that
regular separable potentials in ellipsoidal coordinates, associated
with finite total mass and finite flattening at large radii, lead to
density distributions that share the asymptotic radial behavior (in
general modulated by angular dependence)
\[
\rho\propto {1\over r^4},\quad r\to\infty,
\label{Rna}
\]
with additional properties listed in Point 4 in the Introduction.

We now use a similar approach in MOND. For a finite mass system, the
leading monopole term of the MOND potential follows from
eq.~(\ref{eqmug}), with $\phi\sim\sqrt{GM\az}\ln r$, due to the
$r^{-1}$ decline of the MOND acceleration at large distances (where
the field is weak, and so the system is actually dMOND). Therefore, if
we ask for separability we are now forced to assume
\[
F(\tau)= -{\tau^2 h(\tau)\ln\tau\over 2},
\quad h(\tau)\sim 1\quad {\rm for}\; \tau\to\infty,
\label{Fma}
\]
where the coefficient $1/2$ takes into account the relation
$\lambda\sim r^2$.  Again, the dimensional coefficient $\sqrt{GM\az}$
is set equal to 1: in the density profile obtained by the application
of the MOND operator, due to its non-linearity, the resulting
coefficient is $GM\az$.
\begin{figure*}
\centerline{
\psfig{file=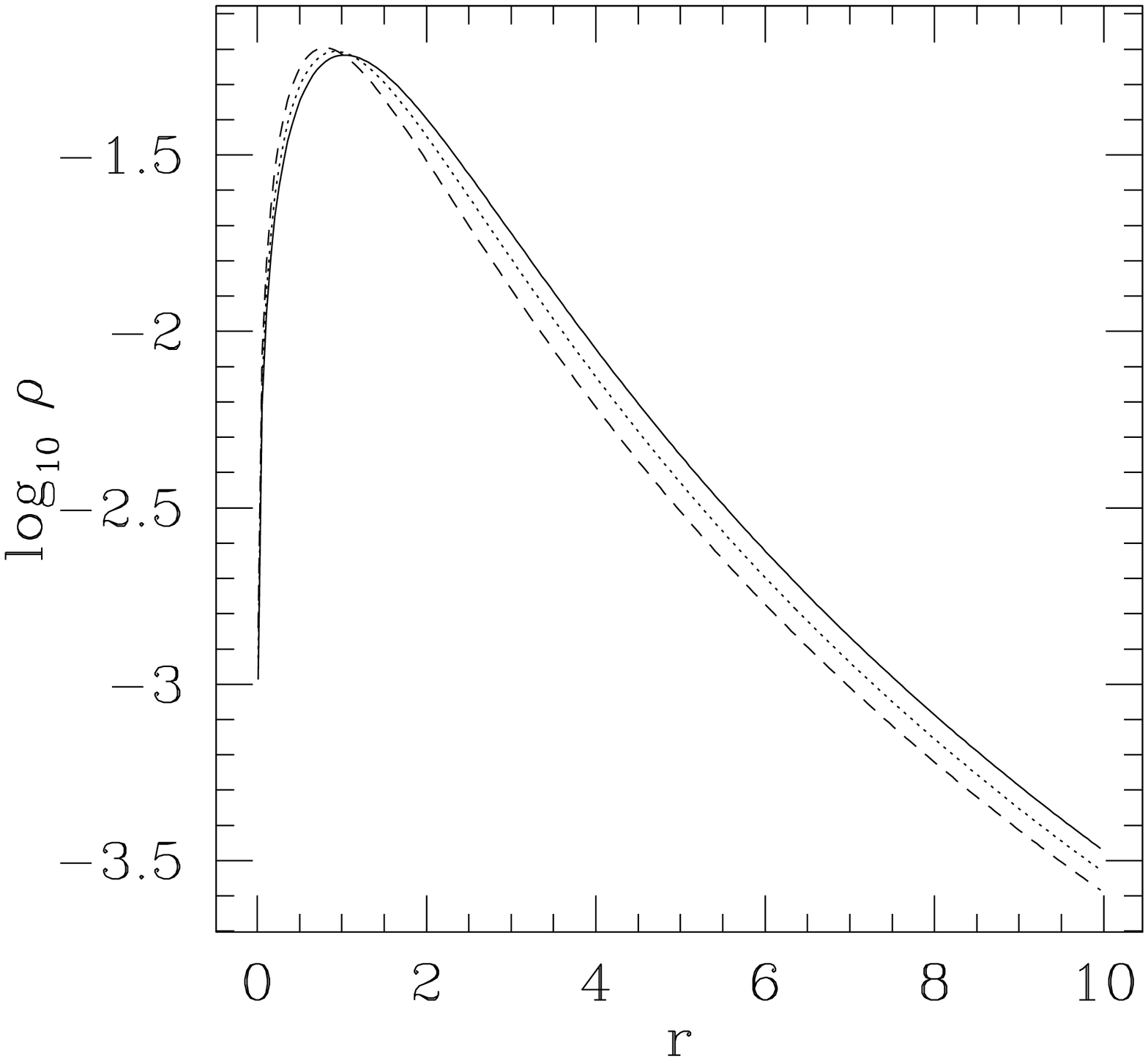,width=0.5\hsize}
\psfig{file=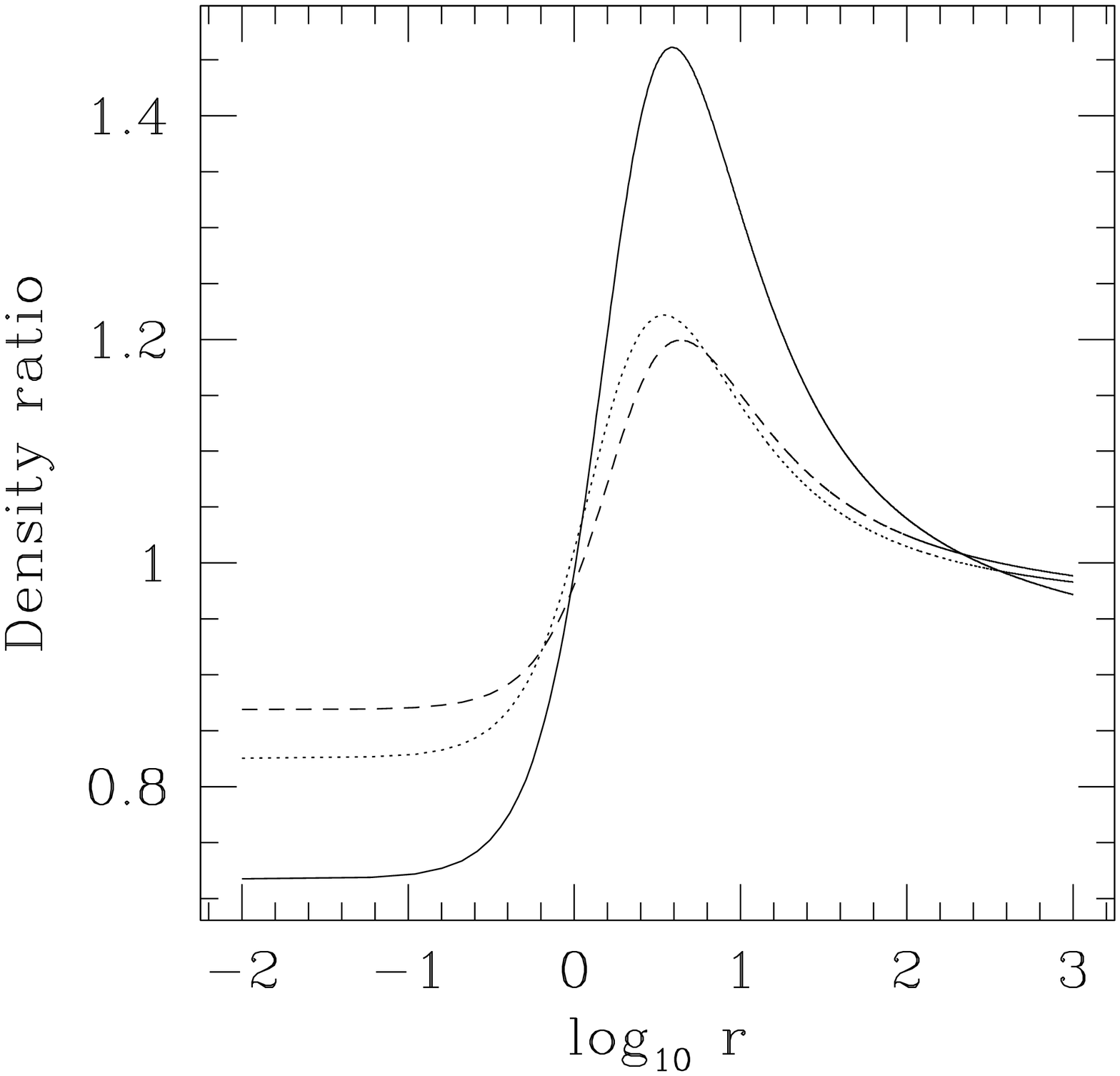,width=0.5\hsize}
}
\caption{{\it Left panel}: density profiles (normalized to $M/4\pi$,
 where $M$ is the total mass of the model) along the Cartesian
 coordinate axes, at radial distance $r$ from the center ($x$: solid,
 $y$: dotted, and $z$: dashed) for a separable model with
 $F(\tau)=-\tau^2\ln(\tau)/2$ and $\alpha=-3$, $\beta=-2$, and
 $\gamma=-1$.  Note how the density vanishes at the center. {\it
   Right panel}: Density ratios $\rho(x,0,0)/\rho(0,0,z)$,
 $\rho(0,y,0)/\rho(0,0,z)$, and $\rho(x,0,0)/\rho(0,y,0)$ as a
 function of distance from the origin, i.e.  $x=y=z=r$, where $\rho$  is
 intended expressed in Cartesian coordinates. Lines are, in
 order, solid, dotted, and dashed, respectively.}
\end{figure*}

As in the Newtonian case, the detailed behavior at infinity of the
shape function $h(\tau)$ determines the radial profile of the density
distribution (note that this is not quite the same as the
three-dimensional shape, which is also a function of the angular
coordinates).  In fact, also the first and second derivatives of $h$
are involved in the computation of $\rho$ and, as is well known,
asymptotic properties of functions in general are not shared by their
derivatives\footnote{An elementary example of a function asymptotic to
  1 with arbitrarily large derivative for $x\to\infty$ is
  $1+x^{-1}\sin x^3$. For a discussion of the possible issues involved
  in the differentiation of asymptotic relations, see e.g. Bender \&
  Orszag (1978).}.  In particular, an arbitrary choice of $h$ can lead
to a system with negative densities or infinite total mass, in
contrast with the hypothesis behind eq.~(\ref{Fma}). For this reason,
the convergence of the total mass must be checked for any specific
choice of $h(\tau)$. In order to carry out a sufficiently general
analysis (encompassing a large fraction of the cases arising in
practical situations), here we restrict to shape functions $h$ with a
Frobenius expansion for $\tau\to\infty$, i.e. we assume
\[
h(\tau)=1 +{A\over\tau^{\epsilon}} + {B\over\tau^{1+\epsilon}}+
{\cal O}\left({1\over\tau^{2+\epsilon}}\right)
\quad {\rm for}\; \tau\to\infty,
\label{eq:hlau}
\]
with $\epsilon\geq 0$. Note that eq.~(\ref{Fma}) 
imposes $A=0$ for $\epsilon=0$, while for
$\epsilon$ integer we are actually dealing with a regular function at
infinity. 
\begin{figure*}
\centerline{
\psfig{file=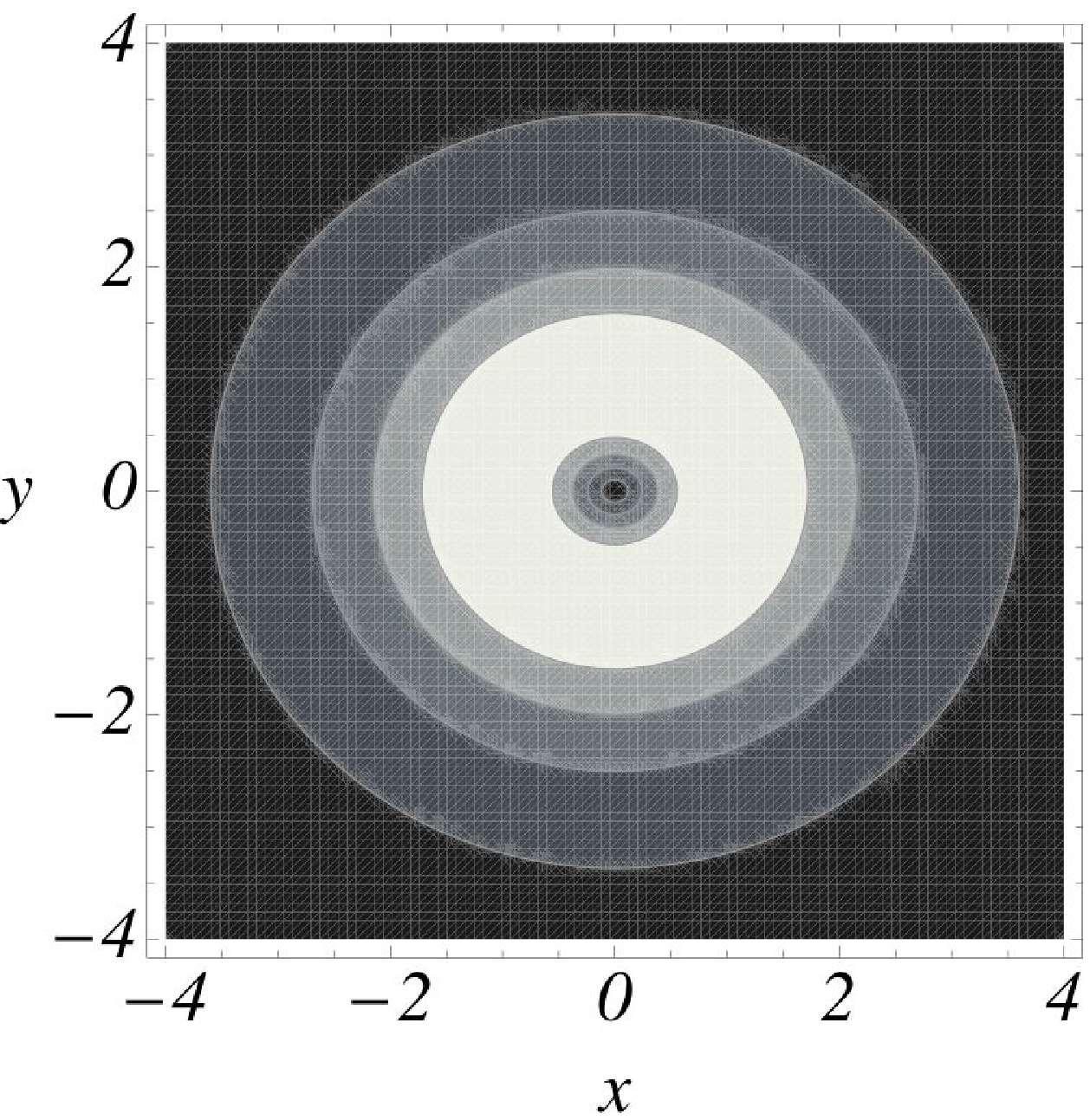,width=0.32\hsize}
\psfig{file=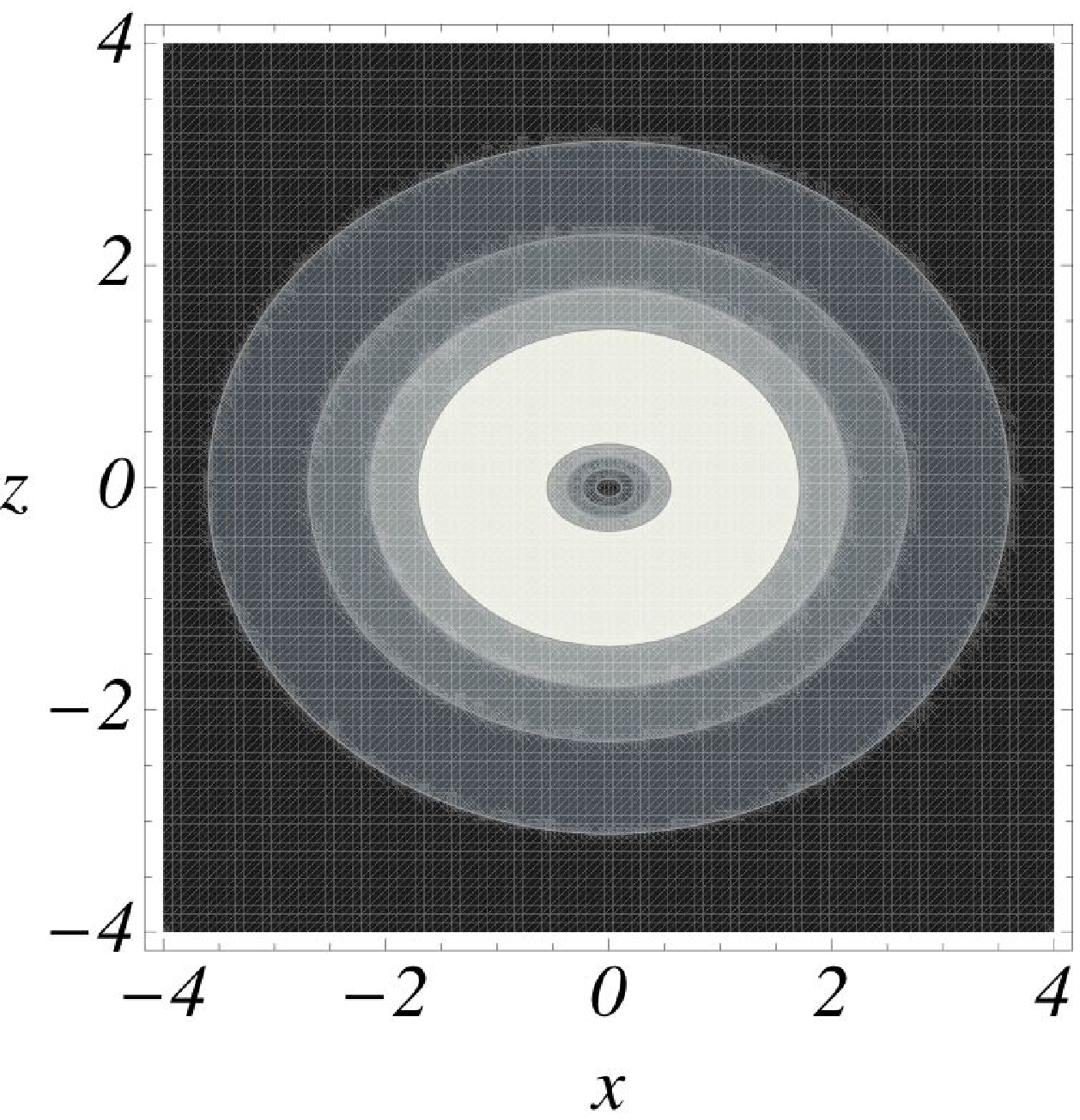,width=0.32\hsize}
\psfig{file=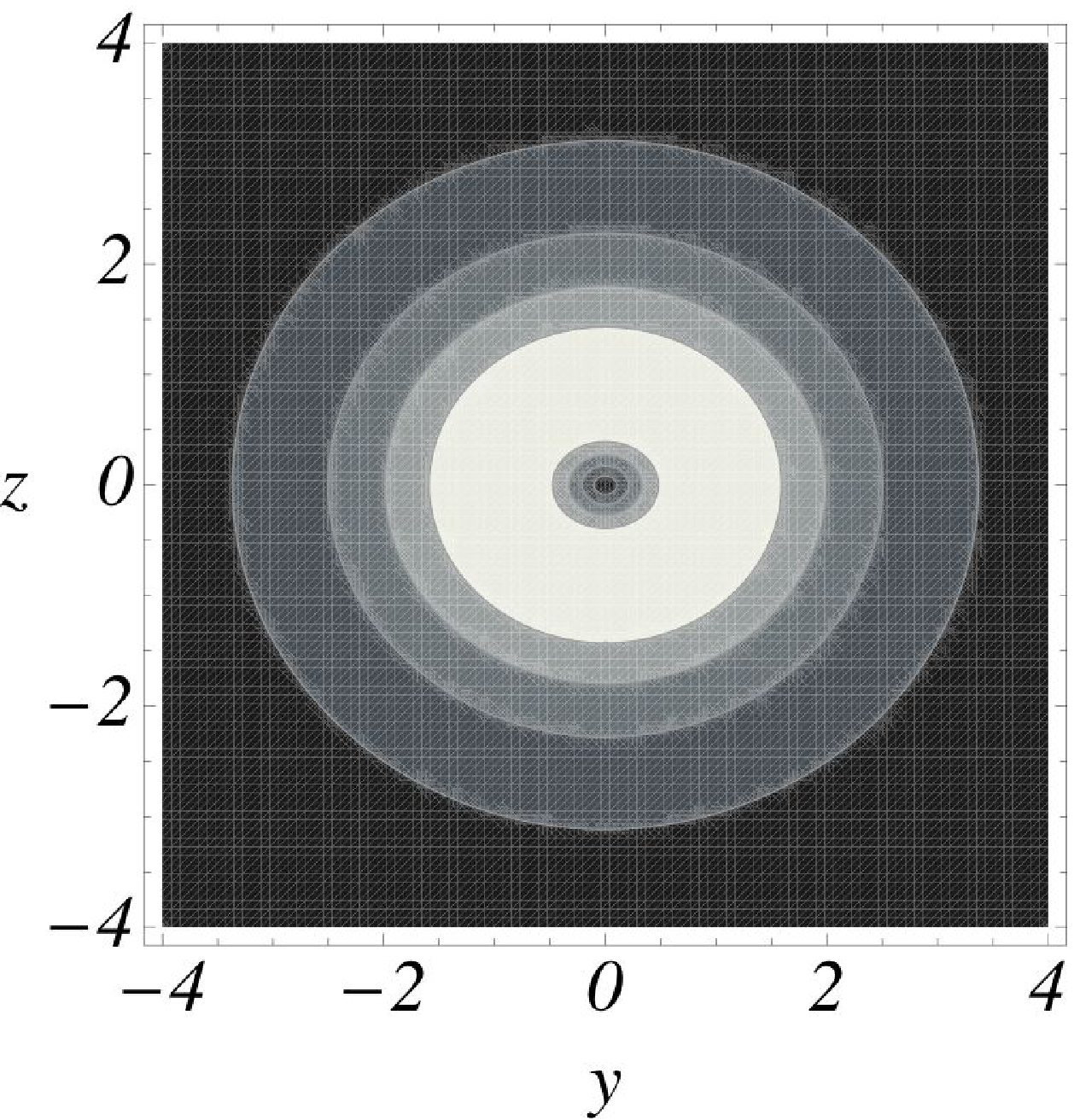,width=0.32\hsize}
}
\caption{Isodensity contours in the three Cartesian coordinate planes
  for the reference model, given by $F(\tau)=-\tau^2\ln(\tau)/2$ (see
  Fig.~1). Darker gray corresponds to lower values of the density.
  From the figure it is apparent how the density decreases near the
  center and at large radii, so that the model presents an ellipsoidal
  corona of higher density. Note also how the $z$ axis corresponds to
  the shortest axis of the density distribution in the radial
    interval shown.}
\end{figure*}

We begin with the regular case $\epsilon =1$.  The computation of the
dMOND operator does not pose special difficulties in the regular case,
and for $\lambda\to\infty$ one finds
\begin{eqnarray}
\cases{
\displaystyle{\nabla^2\phi\sim
{1\over\lambda}-{2(\alpha+\beta+\gamma+\mu+\nu -A/2)
\ln \lambda\over\lambda^2}+{\cal O}(\lambda^{-2}),}\cr
\displaystyle{\Vert\nabla\phi\Vert^2\sim {1\over\lambda}-{2(A+\mu+\nu)\ln
 \lambda\over\lambda^2}+{\cal O}(\lambda^{-2}),}\cr
\displaystyle{\Dphi\Vert\nabla\phi\Vert^2
\sim -{2\over\lambda^2}+{10(A+\mu+\nu)\ln \lambda\over\lambda^3}+{\cal O}(\lambda^{-3}),}
}
\label{eq:coefinf}
\end{eqnarray}
so that, after restoring the dimensional factor, we obtain for $\lambda\to\infty$
\[
\rho\sim {M\over 4\pi}
{(\mu+\nu +4A -2\alpha-2\beta-2\gamma)\ln\lambda\over \lambda^{5/2}}
+{\cal O}\left(\lambda^{-5/2}\right).
\label{Rma}
\]
A few important points should be noted. First, only the leading 
coefficient $A$ appears, while all the higher order coefficients do
not affect the leading term of the density expansion: at infinity,
systems with $A=0$ are indistinguishable from systems with constant
$h=1$. Second, the density distribution at large radii is not spherically
symmetric, a consequence of the exact cancellation of the leading
terms in eq.~(\ref{eq:coefinf}) when combined in
eq.~(\ref{dMONDell}). Third, at large radii the density is nowhere
negative for $A\geq (2\alpha+3\beta+3\gamma)/4$: this positivity
result is not expected a priori, especially when considering the
non-linear nature of the $p$-Laplacian.  Finally, the radial behavior
of the density at large radii is proportional to $\ln(r)/r^5$.
Curiously, the light distribution of elliptical galaxies in their
external regions seems to be described better by the $1/r^4$ profile
(e.g., see Jaffe 1983, Bertin \& Stiavelli 1984, 1989; Hernquist 1990,
Dehnen 1993, Tremaine et al. 1994, see also Bertin \& Stiavelli 1989),
characteristic of the regular Newtonian separable case.

The discussion above concludes the case of an $h$ function with a
regular expansion at infinity.  Moving to the case of non-integer
$\epsilon$, with some additional work it can be shown that for
$\epsilon >1$ eq.~(\ref{Rma}) still holds with $A=0$, thus extending
to the irregular cases the result on the effect of higher order terms
obtained for a regular function $h$.  Therefore, we are left with the
irregular case $0<\epsilon<1$, i.e. when the shape function $h(\tau)$
admits a genuine Frobenius expansion at $\infty$.

Lengthy algebra and a careful order balance show that the following
rigorous formulae, extending the validity of those in
eq.~(\ref{eq:coefinf}), hold:
\begin{eqnarray}
\cases{
\displaystyle{
\nabla^2\phi\sim
{1\over\lambda} +{C\over\lambda^{1+\epsilon}}
-{2(\alpha+\beta+\gamma+\mu+\nu)
\ln \lambda\over\lambda^2}+{\cal O}(\lambda^{-2}),}\cr
\displaystyle{\Vert\nabla\phi\Vert^2\sim 
{(1+D/\lambda^{\epsilon})^2\over\lambda}-
{2(\mu+\nu)\ln\lambda\over\lambda^2}+{\cal O}(\lambda^{-2}),}\cr
\displaystyle{\Dphi\Vert\nabla\phi\Vert^2\sim
-{2(1+D/\lambda^{\epsilon})^2(1+E/\lambda^{\epsilon})\over\lambda^2}+
{10(\mu+\nu)\ln \lambda\over\lambda^3}+{\cal O}(\lambda^{-3}),}
}
\label{eq:irresp}
\end{eqnarray}
where
\begin{eqnarray}
\cases{
C=A[1-4\epsilon -\epsilon (1-2\epsilon)\ln\lambda],\cr
D=A(1-\epsilon\ln\lambda),\cr
E=A[1+4\epsilon -\epsilon (1+2\epsilon)\ln\lambda].
}
\end{eqnarray}
For $\epsilon=1$ we recover the results in
eq.~(\ref{eq:coefinf}).  Combining the expansions above in
eq.~(\ref{dMONDell}), and expanding the norm at the denominator, some
additional work finally shows that
\begin{eqnarray}
\rho&\sim&
{M\over 4\pi}{4A\epsilon (\epsilon\ln\lambda -2)(1+C/\lambda^{\epsilon})
\over\lambda^{3/2+\epsilon}}+\cr
&&
{M\over 4\pi}{(\mu+\nu
 -2\alpha-2\beta-2\gamma)\ln\lambda\over\lambda^{5/2}}+
{\cal O}\left(\lambda^{-5/2}\right),
\label{eqinf}
\end{eqnarray}
and again for $\epsilon =1$ the regular case in eq.~(\ref{Rma}) is
recovered.  The density profile at large radii consists of the
Frobenius contribution dependent on $A$, $\epsilon$, and $\lambda$,
and in the regular part, independent of $\epsilon$. The first
component becomes spherical at large radii, at variance with the
regular component, dependent also on the $\mu$ and $\nu$ coordinates.
In addition, the spherical component is dominant for $0 <\epsilon <1$,
being asymptotic to $4A\epsilon^2\ln
(\lambda)/\lambda^{3/2+\epsilon}$.  Therefore, when $0<\epsilon<1$ the
density at large radii is spherical and positive for $A>0$, while for
$\epsilon \geq 1$ it is non-spherical: it is always positive for
$\epsilon >1$, and for $\epsilon =1$ positivity is assured for $A$
greater than some negative value. We conclude by noticing that the
choice of the potential (\ref{Fma}) implicitly assumes a finite total
mass, and in fact this is found in the solution, as
$\rho\propto\ln(r)/r^{3+2\epsilon}$ for $0<\epsilon <1$ and
$\rho\propto\ln(r)/r^5$ for $1\leq\epsilon$.

\section{Explicit cases}

From the general analysis in Section 3 we found that the central
density of MOND models vanishes for regular potentials separable in
ellipsoidal coordinates. We also found that at large radii the
positivity is assured, provided a certain coefficient in the series
expansion of the shape function at infinity is larger than a threshold
value (0 in a genuine Frobenius expansion, and negative in the regular
case, with the specific value given by a linear combination of the
three axial coefficients $\alpha$, $\beta$ and $\gamma$ defining the
ellipsoidal coordinate system).  Therefore, we have at least
indications that everywhere positive triaxial densities with separable
potentials may exist in MOND.

Unfortunately, without the explicit Kuzmin formula, in general the
positivity of the density associated with an assigned $F(\tau)$ can be
checked only numerically. We now show that such cases in fact can be
found routinely.  For illustrative purposes, we start with a
representative dMOND positive separable model, then we illustrate with
a few examples how the shape function $h$ affects the resulting
densities. For simplicity, all the examples presented in this Section
are considered in the dMOND regime. The use of the full MOND equation
would introduce a dependence on total mass, leading to a more
complicated discussion, without adding new information to the present
discussion.

The reference model is perhaps the simplest possible, and it is
obtained by using $F(\tau)=-\tau^2\ln(\tau)/2$ in eq.~(\ref{phisep}),
i.e., we just fix $h=1$ in eq.~(\ref{Fma}). We also assume
$\alpha=-3$, $\beta=-2$, and $\gamma=-1$, but we stress that global
positivity has been found for all the explored values of the three
parameters, even in the cases characterized by very different values
of the axial parameters.  The relations between ellipsoidal and
Cartesian coordinates needed for the plots of the isodensity contours
in the three coordinate planes are reported in Appendix A3.

In the left panel of Fig.~1 we show the density profiles
$\rho(x,0,0)$, $\rho(0,y,0)$, and $\rho(0,0,z)$ of the reference
model, where it is intended that now the density is expressed in
Cartesian coordinates, and $x=y=z=r$, where $r$ is the radial distance
from the origin.  It is apparent how the density vanishes at the
center, then reaches a peak, and finally declines again, in accordance
with the previous asymptotic analysis.  The model is not spherical, as
can be seen from Fig.~2, where we present the density cross-sections
(in the central regions) in the three Cartesian coordinate planes:
dark grays correspond to low density values, while light grays are the
density peaks. All the main features of Fig.~1 can be easily
recognized, and in particular the density depression in the inner
regions: overall the density of the reference model is characterized
by a very nice ellipsoidal shape.  In practice, the resulting density
distribution looks similar to an heterogeneous ellipsoid with a
non-monotonic density stratification.  The radial trend of the density
shape is quantified, on a much larger radial interval, in the right
panel of Fig.~1, where the density ratios $\rho(x,0,0)/\rho(0,0,z)$,
$\rho(0,y,0)/\rho(0,0,z)$, and $\rho(x,0,0)/\rho(0,y,0)$ are
represented by the solid, dotted, and dashed lines, respectively, for
$x=y=z=r$.  Note that in general the density ratios
$\rho(r,0,0)/\rho(0,0,r)$, and $\rho(0,r,0)/\rho(0,0,r)$, for a
density distribution flattened along the $z$ direction, are $\geq 1$
($\leq 1$) if the density is decreasing (increasing) with $r$.  A
visual inspection of Fig.~2 then explains the behavior of the density
ratios up to $r\approx 100$: in these regions the $z$ axis (the long
axis of the coordinate ellipsoids) corresponds to the ``short'' axis
of the density distribution, thus confirming that this property
usually holds not only in Newtonian separable systems, but also in
MOND. We also note the interesting occurrence of a double switch
between the intermediate and the long axis. However, for large $r$,
the density ratios drop again below $1$, i.e., the density
distribution in these regions (not shown in Fig.~2) becomes elongated
along the $z$ axis: it has been verified that this non-sphericity is
in perfect agreement with eq.~(\ref{Rma}) evaluated for
$A=0$. Therefore, this model represents a counterexample (in MOND) for
the Eddington (1915) conjecture, fully discussed in by de Zeeuw et
al.~(1986, Section 4 therein).

\begin{figure*}
\centerline{
\psfig{file=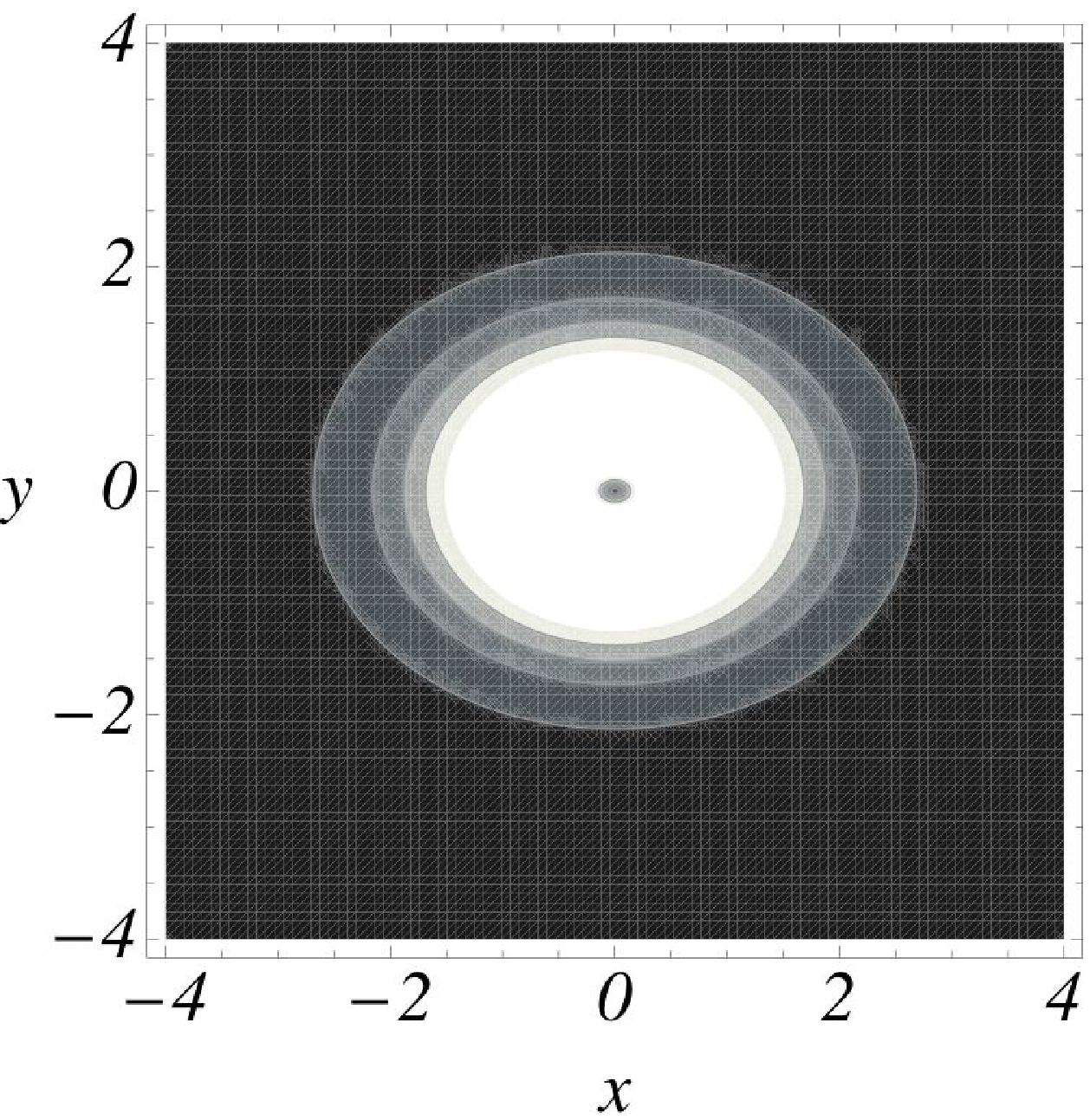,width=0.31\hsize}
\psfig{file=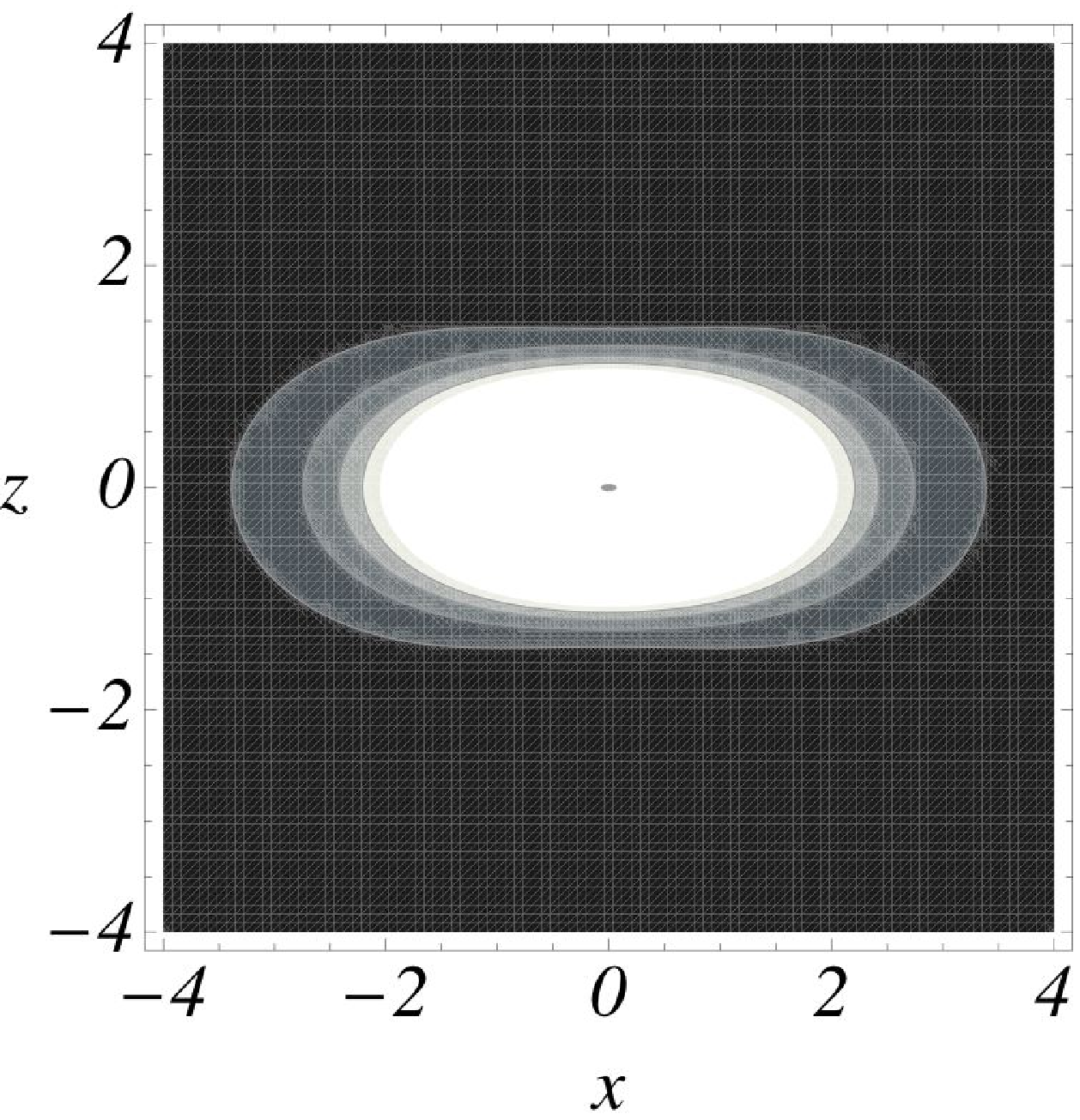,width=0.31\hsize}
\psfig{file=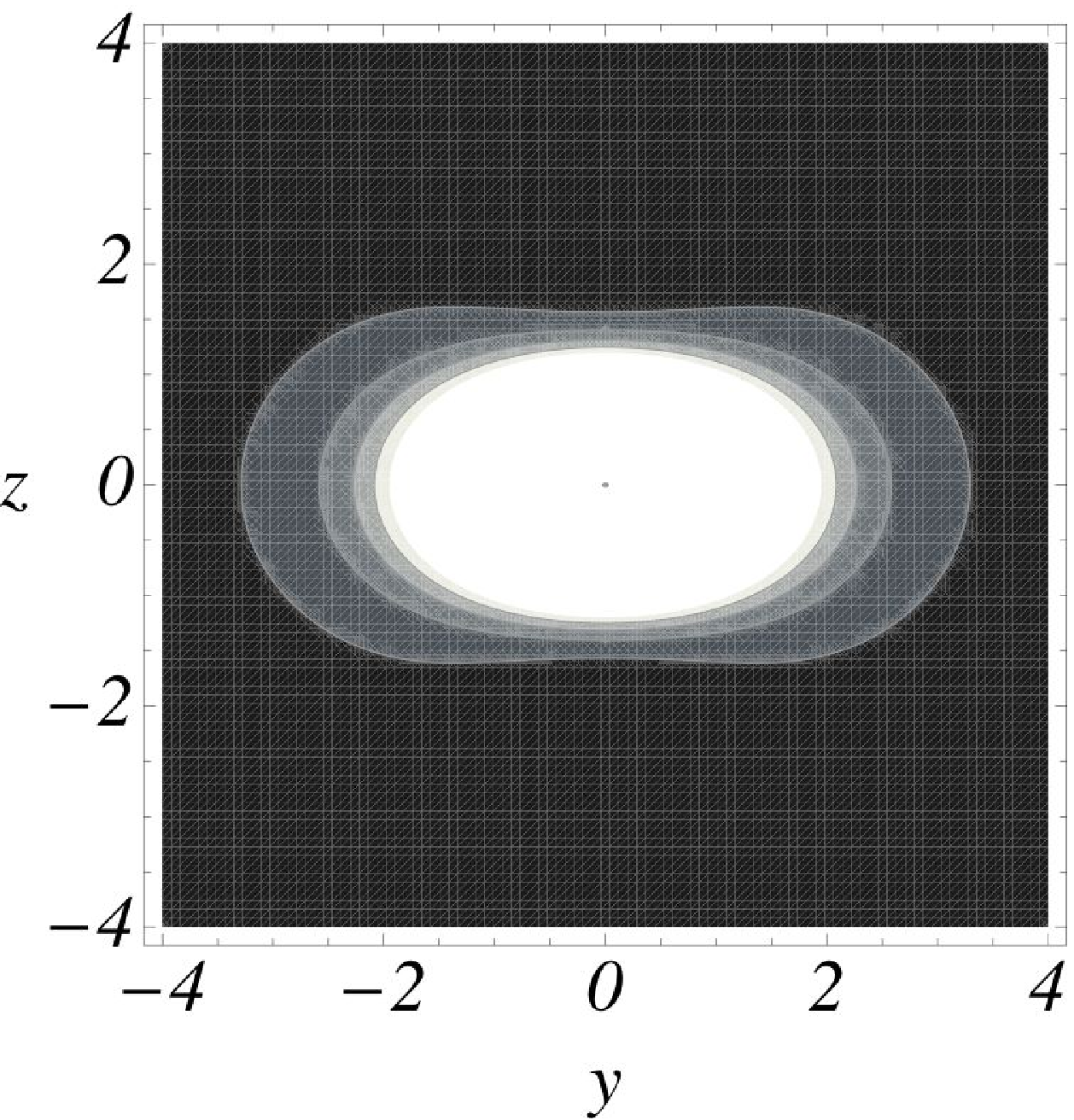,width=0.31\hsize}
}
\centerline{
\psfig{file=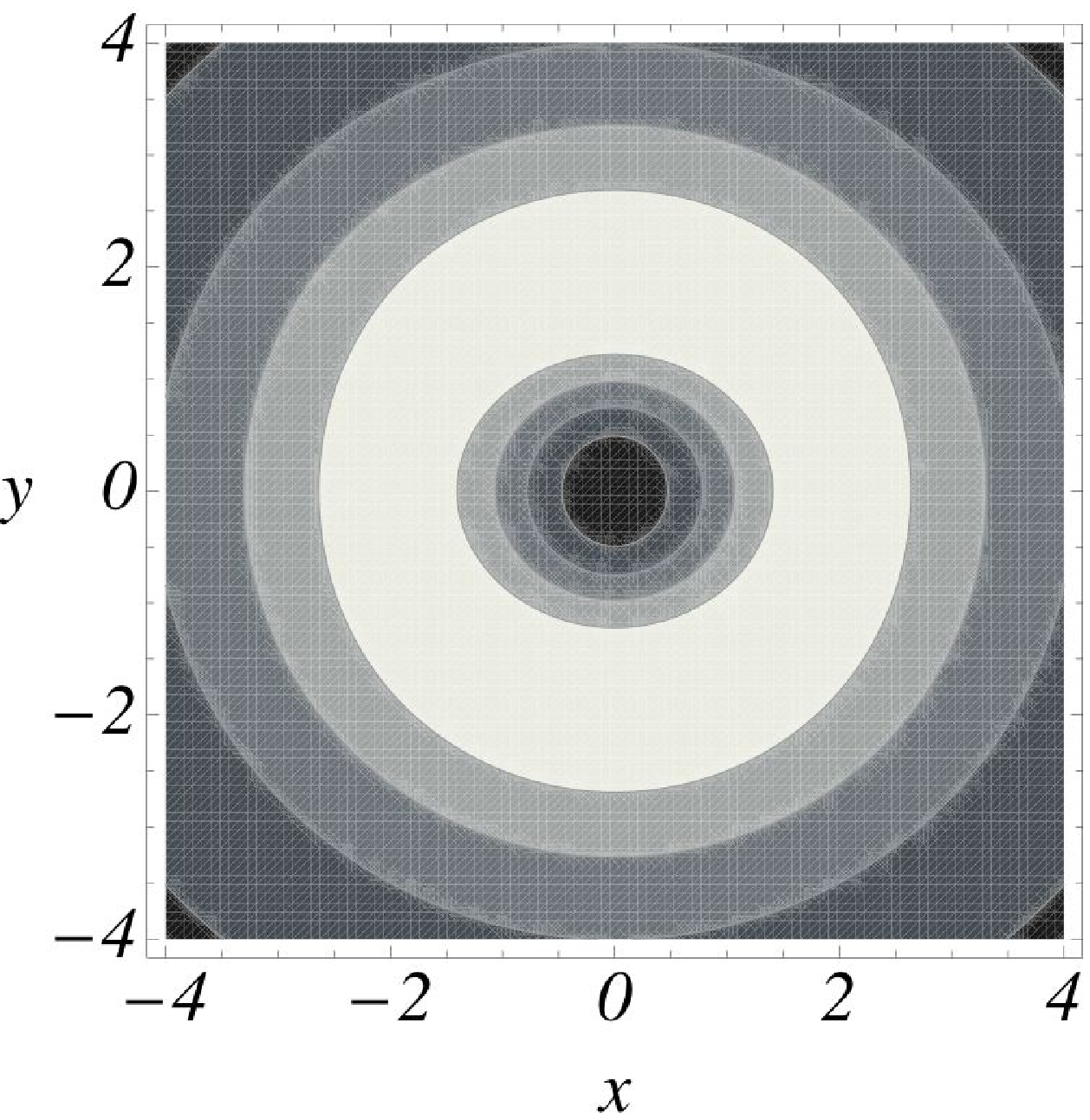,width=0.31\hsize}
\psfig{file=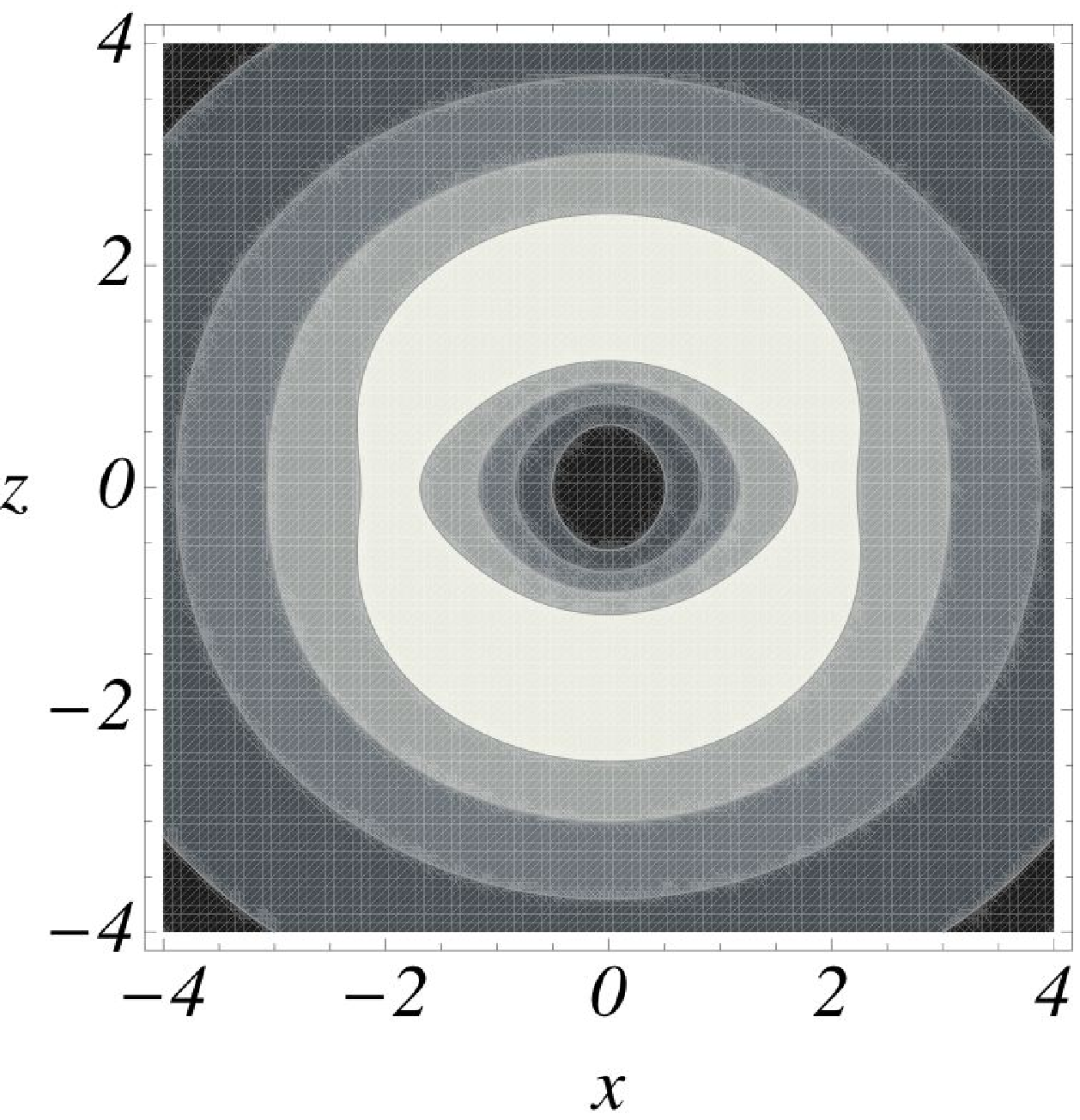,width=0.31\hsize}
\psfig{file=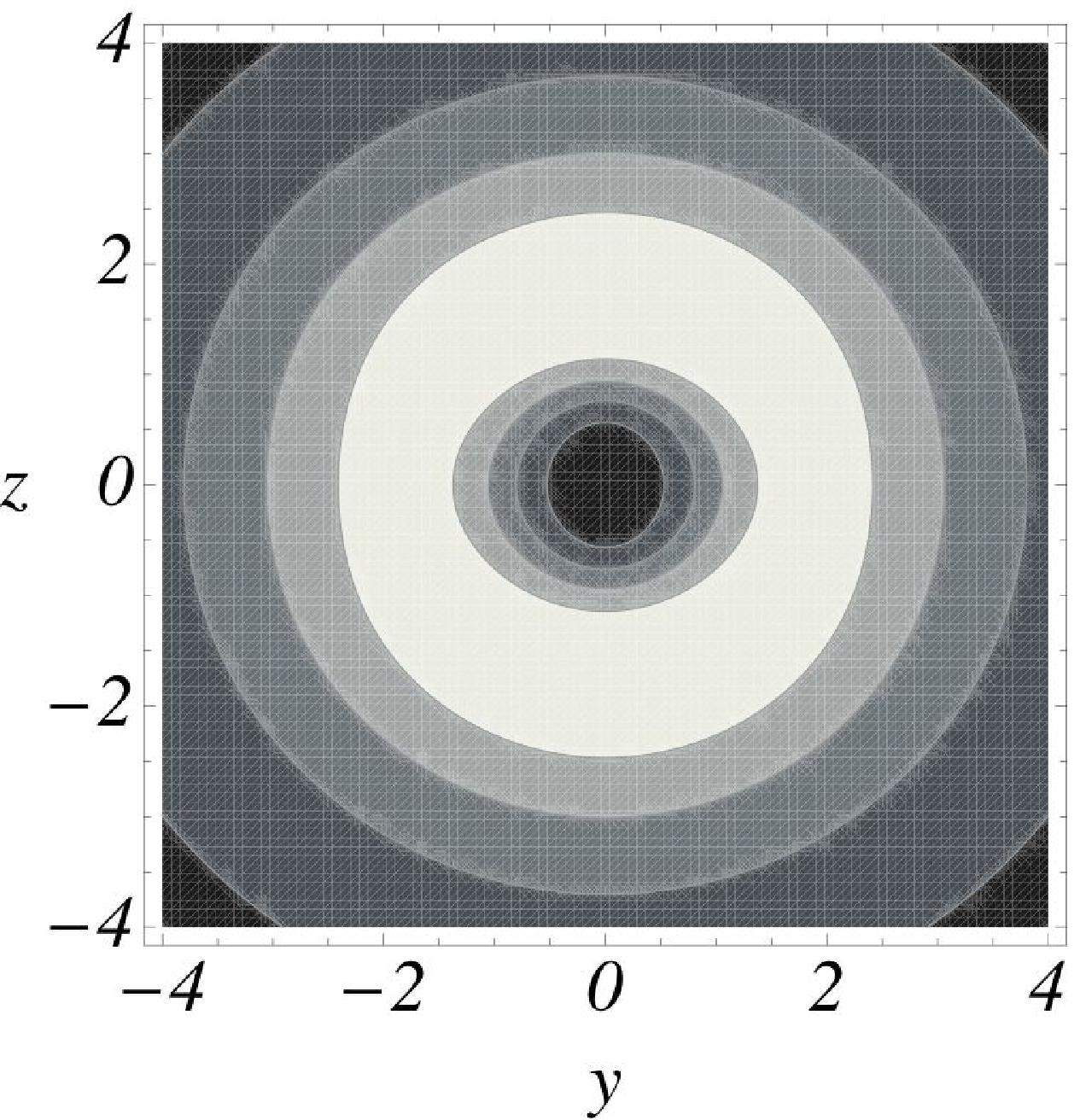,width=0.31\hsize}
}
\centerline{
\psfig{file=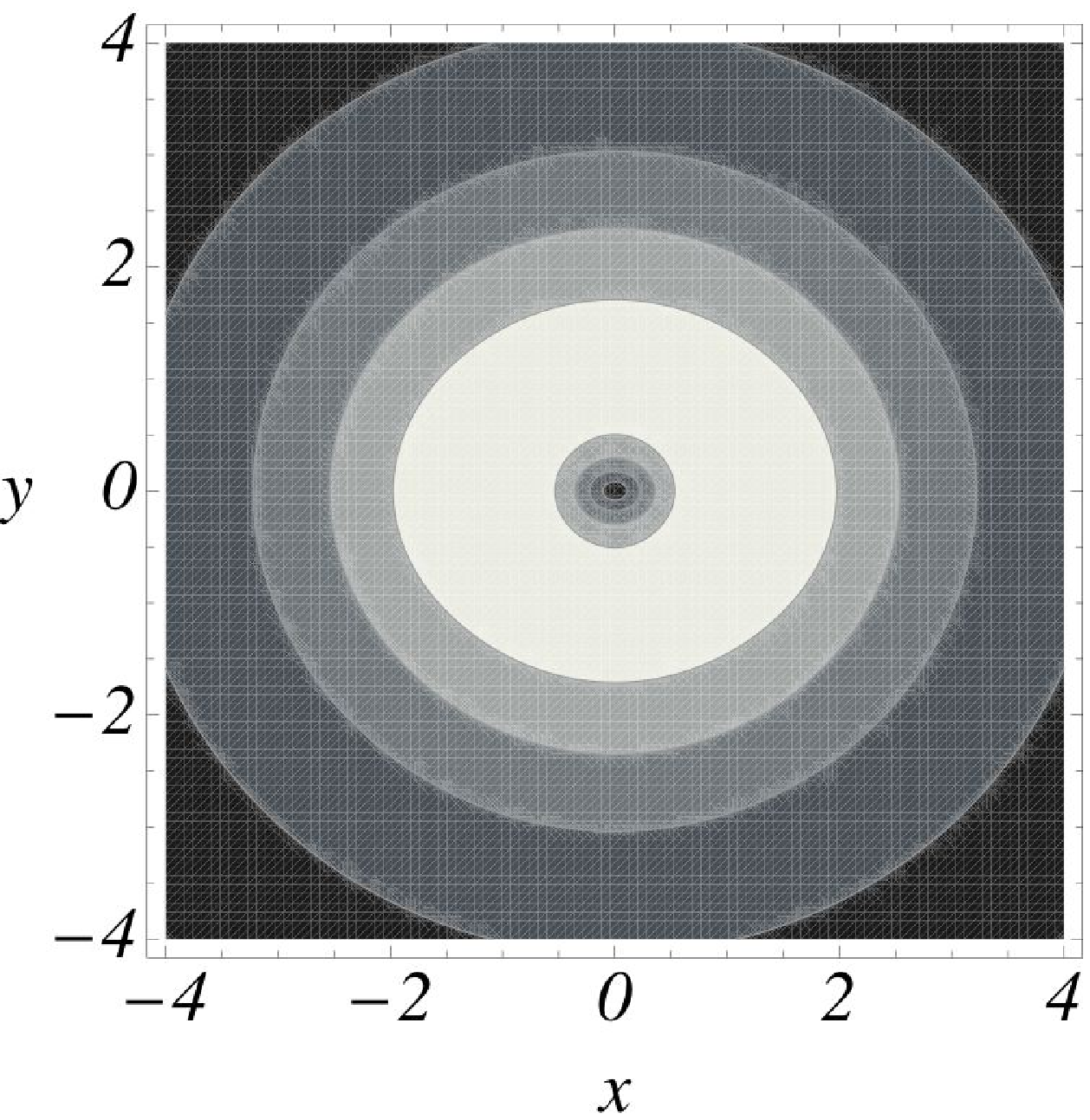,width=0.31\hsize}
\psfig{file=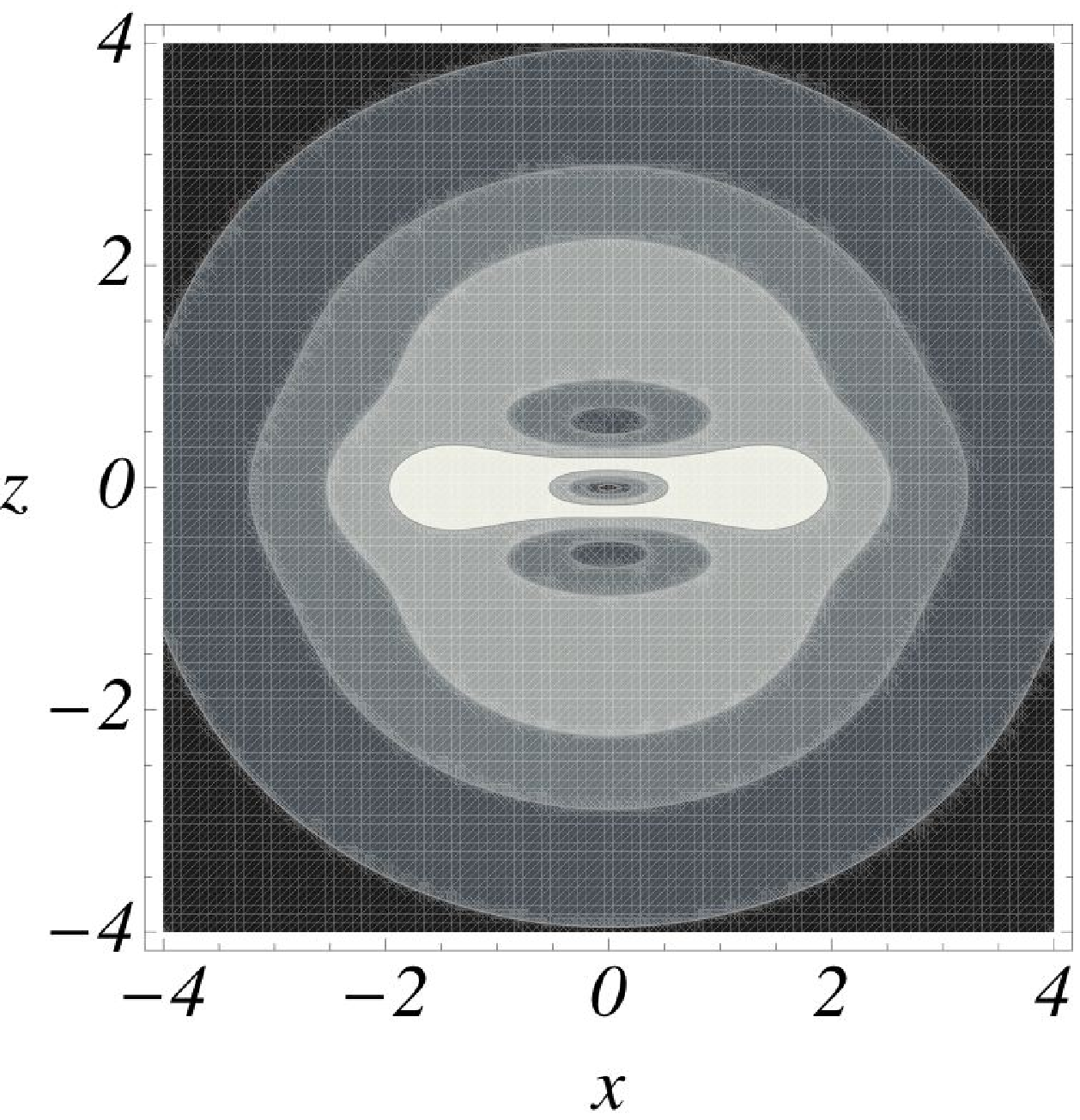,width=0.31\hsize}
\psfig{file=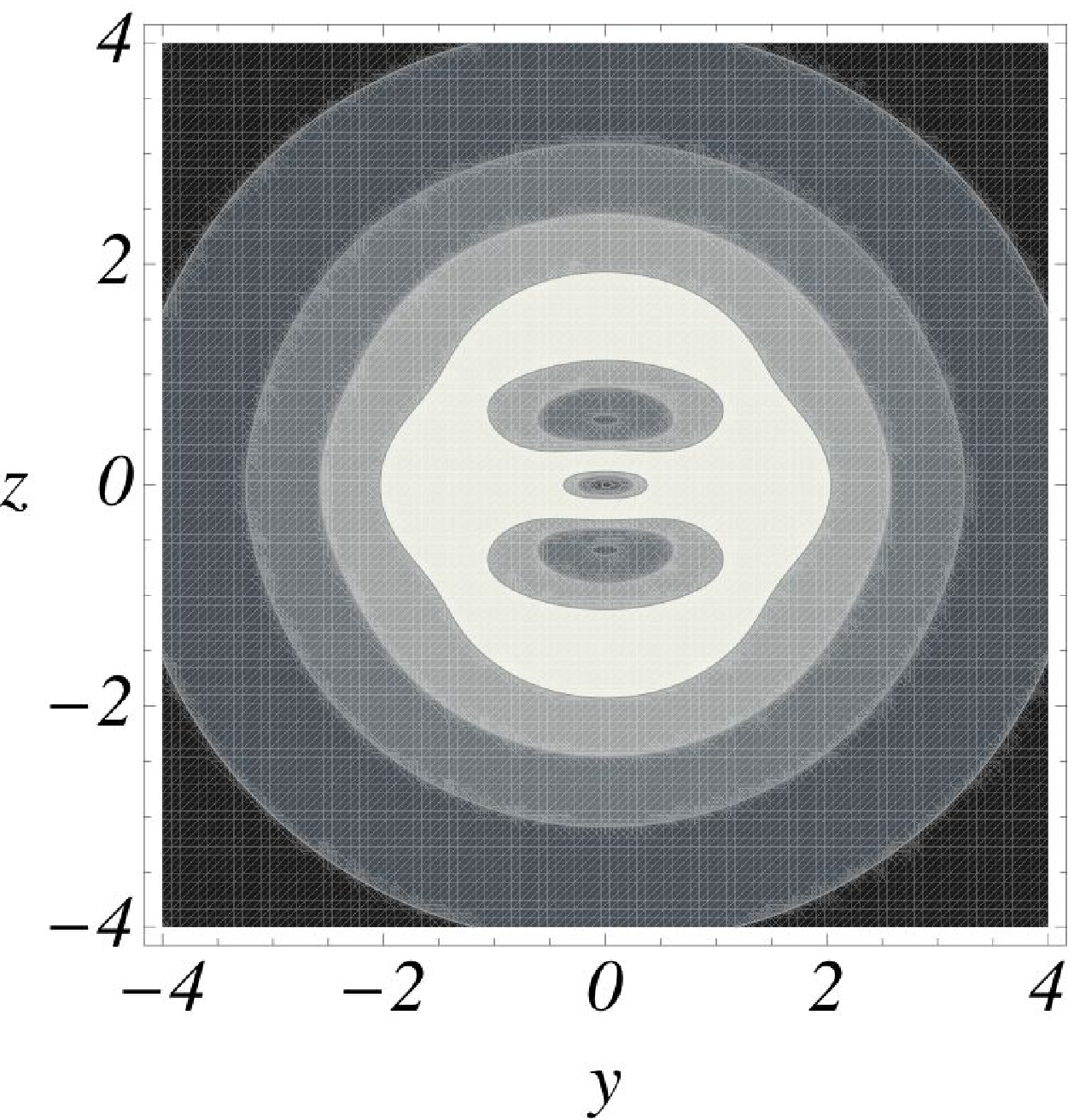,width=0.31\hsize}
}
\centerline{
\psfig{file=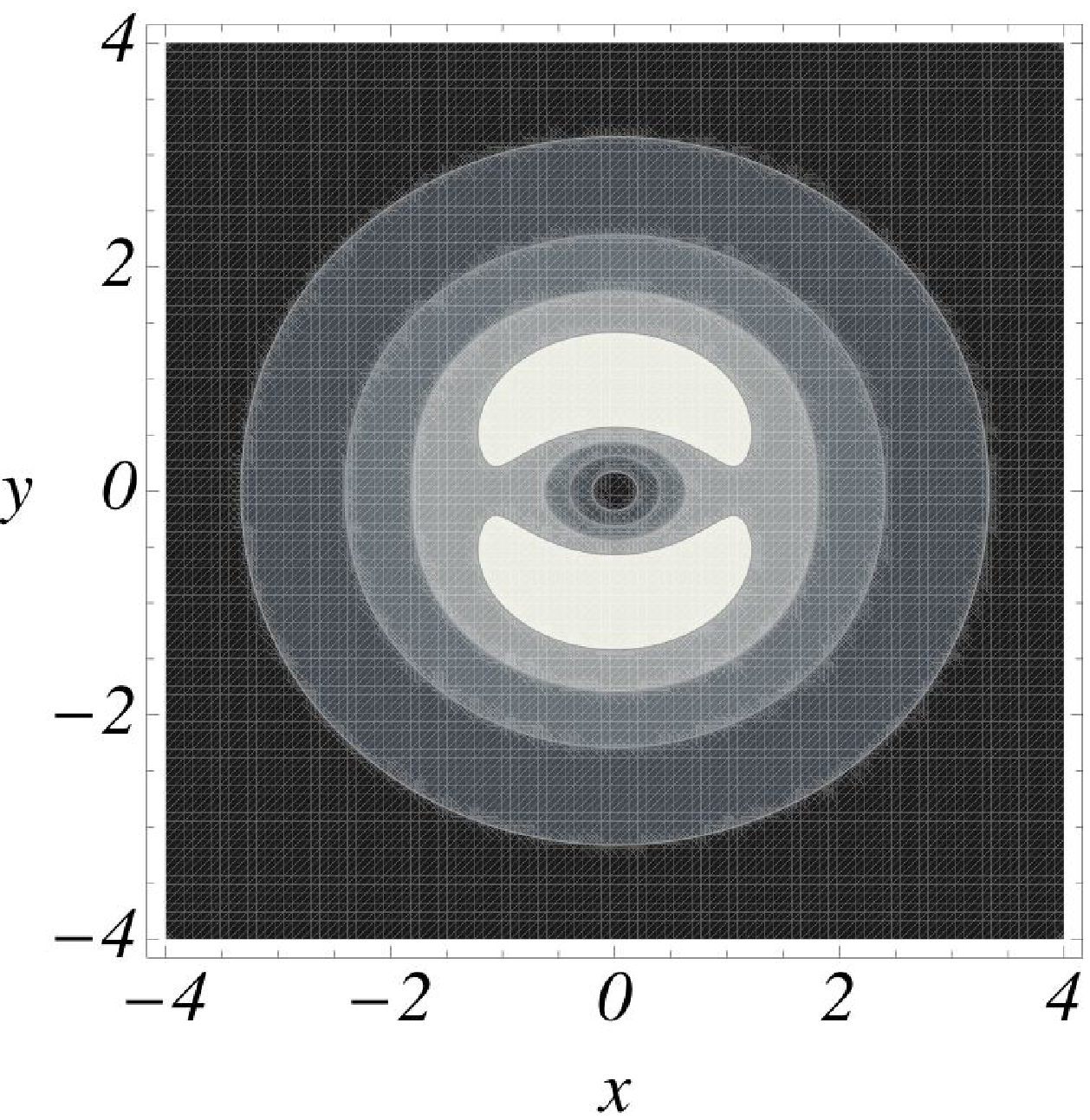,width=0.31\hsize}
\psfig{file=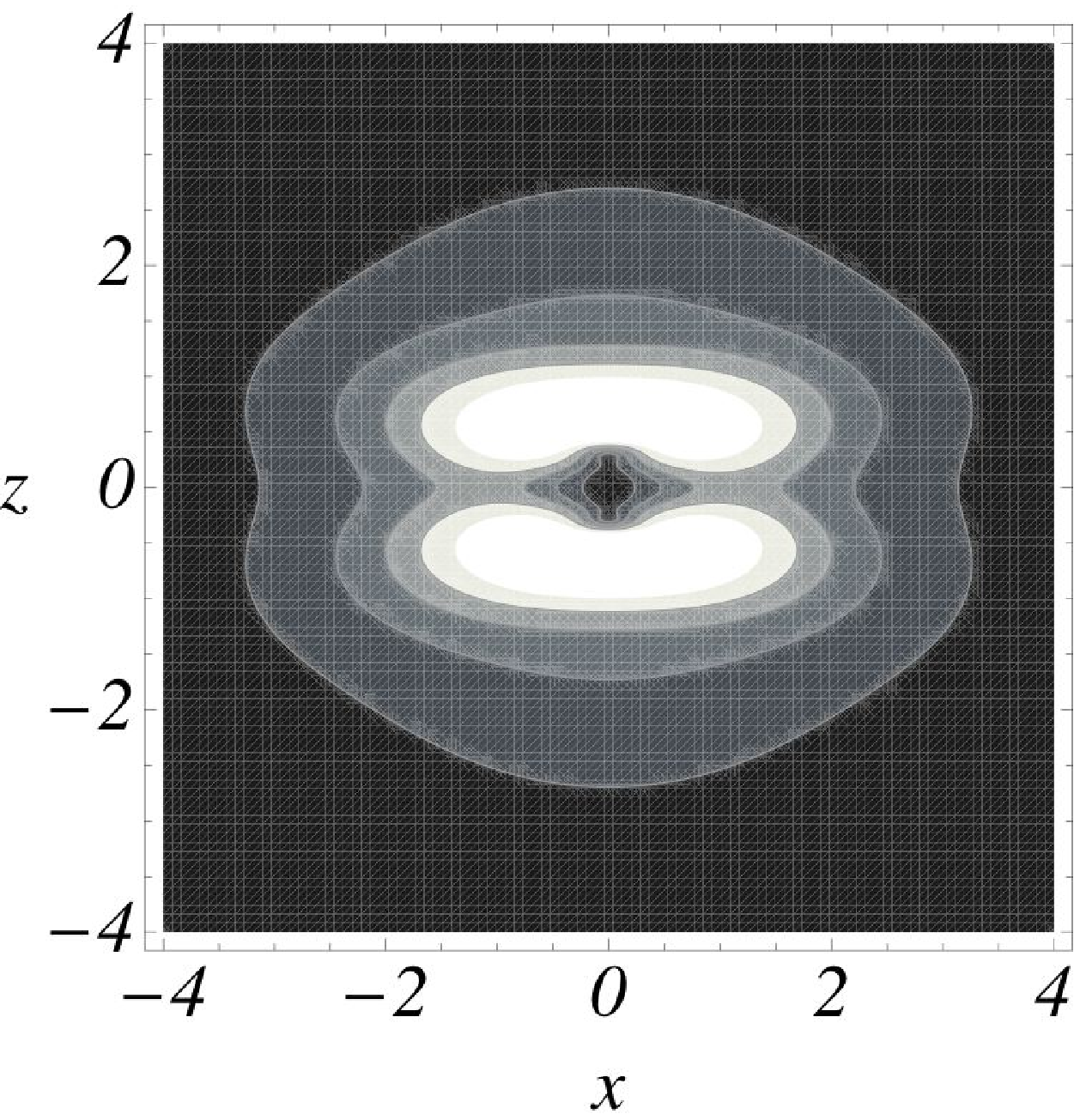,width=0.31\hsize}
\psfig{file=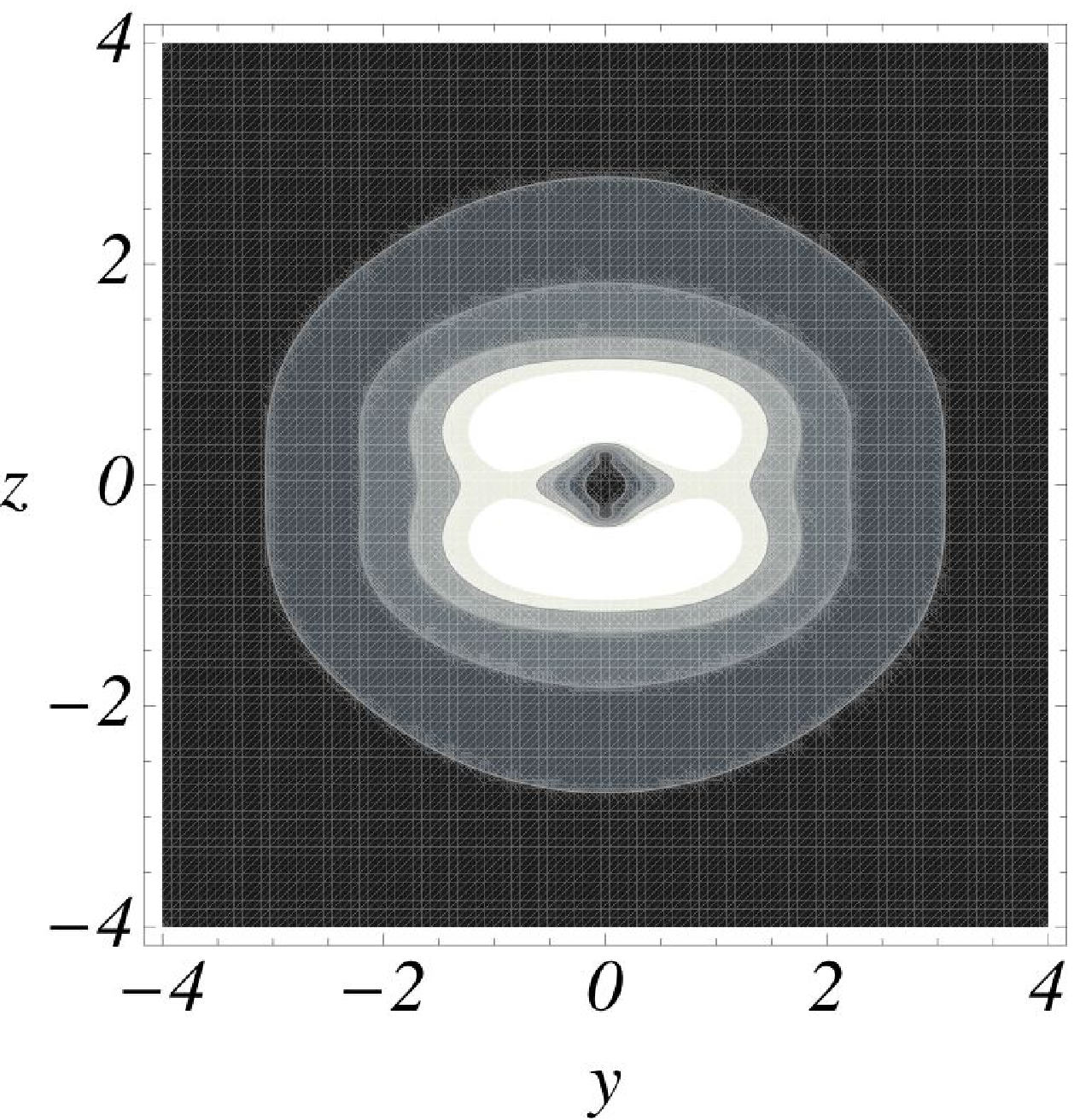,width=0.31\hsize}
}
\caption{Isodensity contours in the coordinate planes for different
  choices of the shape function $h$ in eq.~(\ref{Fma}), and for
  $\alpha=-3$, $\beta=-2$, and $\gamma=-1$. From top to bottom:
  $h(\tau)=1+4/\tau+11/\tau^2$, $1+3/\tau$, $1+2/\tau+1/2\tau^5$, and
  $1+1/\tau^3 -\sqrt{2}/\tau^6$. Dark grays correspond to lower
  density values.}
\end{figure*}

We stress that the remarkable ellipsoid-like density distribution of
the reference model (preserved also for significantly different values
of the axial parameters $\alpha$, $\beta$, and $\gamma$) is not a
general property of MOND models with separable potentials.  In models
where we allow for a non-constant shape function, the resulting
(positive) densities are quite peculiar, in some cases with
high-density, detached lobes along the $z$ axis or, in other cases, by
the presence of curious low-density regions. Of course, in accordance
with the asymptotic analysis, the central density of all these models
still vanishes. In Fig.~3 we show a suite of such densities obtained
for different choices of the function $h$, listed in the caption, and
for the same values $\alpha=-3$, $\beta=-2$, and $\gamma=-1$ of the
reference model in Fig.~2. In particular, moving from the top to the
bottom rows the coefficient $A$ in eq.~(27) decreases, and the
corresponding densities become more and more complicated. This is not
surprising, because the coefficient $A$ approaches the positivity
limit for the density at large radii discussed after
eq.~(\ref{Rma}). Finally, the comparison of the model in the last row
with the reference model in Fig.~2 shows how higher-order terms in the
expansion of $h$ may affect the density, as $A=0$ in both cases.

\section{The $V_1$ family of de Zeeuw \& Pfenniger (1988)}

As discussed in the Introduction, this paper focuses on MOND triaxial
models with the potential separable in ellipsoidal
coordinates. Separable models are a very special subset of
triaxial models, and so it is of some interest to see what properties
of separable models are in fact shared by more general triaxial
models, and what are the main properties of the density distributions
associated to some MOND (non separable) potentials in
ellipsoidal coordinates. Here we restrict for sake of simplicity to
the important $V_1$ family 
\[
\phi=F(\lambda)+F(\mu)+F(\nu),
\label{phi1}
\]
introduced and fully discussed in ZP88. $V_1$ potentials are relevant
here because in Newtonian gravity they obey the Kuzmin theorem; in
addition, altough non-separable, they are algebraically simpler than
separable potentials (which comprise the family $V_2$ and can be
derived from $V_1$ by application of a linear operator [ZP88]).  Based
on the results obtained for separable models, it is reasonable to
expect that also $V_1$ potentials in MOND satisfy the Kuzmin property.

In fact, we now show that the restriction to the $z$-axis of the
r.h.s. of eq.~(\ref{dMONDell}) with the potential (\ref{phi1}),
reduces in the three intervals of $\tau$ in eq.~(\ref{eq:zaxis}) to
the same second-order ODE. This proves that the Kuzmin property holds
also for $V_1$ potentials in dMOND regime (i.e., for the
$p$-Laplacian).  The Laplace operator satisfies the Kuzmin property,
so there is nothing to prove, and its expression along the $z$-axis is
given in Sect.~3.1 of ZP88.  An explicit computation then shows that
over the whole $z$-axis
\[
||\nabla\phi ||_z^2=4(\tau+\gamma)F'(\tau)^2
\label{eq:nab1}
\]
and 
\[
[\Dphi||\nabla\phi ||^2]_z=4||\nabla\phi ||_z^2
\left[F'(\tau)+2(\tau+\gamma)F''(\tau)\right].
\label{eq:nab2}
\]
The following second-order ODE along the 
$z$-axis for dMOND $V_1$ systems is finally obtained:
\begin{eqnarray}
\pi G\az\psi(\tau)&=&(\tau+\gamma)^{3/2}|F'(\tau)|\times\cr
 &&                               \Big[
                                   4F''(\tau)+\left({2\over\tau+\gamma}+{1\over\tau+\alpha}+{1\over\tau+\beta}\right)F'(\tau)\cr
&&                                  -{(\gamma -\alpha)F'(-\alpha)\over (\tau+\gamma)(\tau+\alpha)}                               
                                  -{(\gamma -\beta)F'(-\beta)\over (\tau+\gamma)(\tau+\beta)}                               
                                \Big].
\label{eq:ODE1}
\end{eqnarray}
Again, in analogy with the separable case, it is not difficult
to show that the Kuzmin property (and in principle a Kuzmin formula)
also holds for $V_1$ potentials in the full MOND regime.

As expected, eq.~(\ref{eq:ODE1}) is considerably simpler than the
corresponding eq.~(\ref{eq:ODEz}), and some additional classification
and elaboration can be carried out.  In fact, albeit
eq.~(\ref{eq:ODE1}) is still second order, non-homogeneous and
non-linear, the function $F(\tau)$ is missing, so that {\it for
  general $V_1$ systems in dMOND, the $z$-axis ODE can be reduced to a
  non-linear first order equation}, solving for $F'$.  In particular,
in each $\tau$ interval where the sign of $F'$ is constant, the ODE
belongs to the important family of Abel differential equations of the
second kind:
\[
[g_0(x)+g_1(x)y]y'=f_0(x)+f_1(x)y+f_2(x)y^2+f_3(x)y^3,
\label{eq:abelODE}
\]
(e.g., Kamke 1948, Zwillinger 1997)\footnote{Second kind Abel ODEs
  can be always rewritten as first kind Abel ODEs for $w$ with the
  transformation $g_0(x)+g_1(x) y =1/w$.}, a generalization of the
Riccati equations (Ince 1964). Clearly, the problem is complicated by
the fact that the sign of $F'$ is not known a priori: in the following
discussion we assume, for simplicity, that $F'$ does not change sign
for $\tau\geq -\gamma$, and so we set $F'(\tau)=\pm H(\tau)$, with
$H(\tau)\geq 0$. Under this assumption, eq.~(\ref{eq:ODE1})
can be rewritten as  
\[
H\,H'=\pm f_0(\tau)+f_1(\tau)H+f_2(\tau)H^2,
\label{eq:abelODEV1}
\]
so that in our case we have an Abel equation with $g_0=f_3=0$,
$g_1=1$, and
\begin{eqnarray}
\cases{
\displaystyle{f_0={\pi G\az\psi(\tau)\over 4(\tau+\gamma)^{3/2}}},\cr
\displaystyle{f_1={(\gamma -\alpha)H(-\alpha)\over 4 (\tau+\gamma)(\tau+\alpha)}+
      {(\gamma -\beta)H(-\beta)\over 4(\tau+\gamma)(\tau+\beta)}},\cr
\displaystyle{f_2= -{1\over 4}\left({2\over\tau+\gamma}+{1\over\tau+\alpha}+
        {1\over\tau+\beta}\right)}.
}
\end{eqnarray}
The choice of the sign in front of $f_0$ determines (if they exist)
two solutions $H_{\pm}(\tau)\geq 0$, so that the problem is finally
solved for $F'(\tau)=\pm H_{\pm}(\tau)$. Unfortunately, the general
solution of Abel ODEs (first and second kind) is not known, but
several remarkable transformations have been found (e.g., Polyanin \&
Zaitsev 2003).  For example, by setting $H=g(\tau)p(\tau)$ it is
possible to determine $p(\tau)$ so that eq.~(\ref{eq:abelODEV1}) can
be written in the reduced (but not yet canonical) form
\[
g\,g'=R g \pm S,\quad R={f_1\over p},\quad S={f_0\over p^2};
\label{eq:RSabel}
\]
in our case 
\[
p(\tau)={1\over \sqrt{\tau
    +\gamma}|\tau+\alpha|^{1/4}|\tau+\beta|^{1/4}}.
\label{eq:pabel}
\]
The canonical form would be then obtained by requiring
$R(\tau)=1$, through the definition of the new independent variable
$\xi=\int R d\tau$. Unfortunately, in the triaxial case the evaluation of the
integral requires Appell functions, so that the inversion
$\tau=\tau(\xi)$ is impossible in closed form. In the prolate case
($\beta=\gamma$) the integral reduces to a standard
hypergeometric function, and inversion is again impossible, but in the
oblate case ($\beta=\alpha$) 
\begin{eqnarray}
\xi=(\alpha-\gamma)H(-\alpha)\cases{
\displaystyle{\arcsin\sqrt{\gamma+\tau\over\gamma-\alpha},\quad 
                   -\gamma\leq\tau\leq-\alpha;}\cr
\displaystyle{{\pi\over 2}-{\rm arcsinh}\sqrt{\alpha+\tau\over\gamma-\alpha},\quad 
                   -\alpha\leq\tau;}\cr
}
\label{eq:can}
\end{eqnarray}
and so the reduction to the canonical form is possible by using
circular and hyperbolic functions. However, the resulting
Abel equation is still unsolvable.

As the general problem is not solvable, we elaborate on the
possibility to solve a {\it restricted} problem, searching for special
solutions with $H(-\alpha)=H(-\beta)=0$, i.e. with $f_1=0$  in 
eq.~(\ref{eq:abelODEV1}). This is done by writing, in
full generality,
\[
H=k(\tau) (\tau+\alpha)^{2l}(\tau+\beta)^{2m},
\label{Gkdelta}
\]
where $l$ and $m$ are integer numbers $\geq 1$, and $k$ is a regular function at
$\tau=-\alpha$ and $\tau=-\beta$. The parity of the exponents 
forces $k$ to be positive for $\tau\geq -\gamma$, consistently with
the non-negativity of $H$, and eq.~(\ref{eq:abelODEV1}) reduces 
to a linear ODE for $k^2(\tau)$, so that only solutions everywhere
positive are acceptable. Some algebra shows that the exact solution is
\[
H_{\pm}(\tau)=p(\tau)
\sqrt{k_0 \pm
  2\int_{-\gamma}^{\tau}f_0(t)(t+\gamma)\sqrt{|t+\alpha|\,|t+\beta|}dt},
\]
where $k_0$ is the free parameter. Note that the integral under square
root is monotonic increasing ($f_0\geq 0$), so that positivity is
assured when $k_0\geq 0$ and the sign $+$ is adopted. If the integral
is convergent for $\tau\to\infty$, then also the sign $-$ can be
adopted, provided $k_0$ is larger than the value of the integral at
infinity. In all cases, the square root cannot vanish for $\tau
=\tau_0 >-\gamma$, as monotonicity of the integral then would produce
negative values of $k^2$ for $\tau >\tau_0$.  It follows that $H$
diverges at $\tau=-\alpha$ and $-\beta$ as $p(\tau)$ (independently of
the value of $l$ and $m$), against the assumption, and reduced
solutions with $f_1$ do not exist.  A simple argument shows that this
negative it is to be expected. In fact, if one is allowed to fix
$H(-\alpha)=H(-\beta)=0$ in eq.~(\ref{eq:abelODEV1}), then the
resulting non-homogeneous ODE becomes {\it first-order linear} for
$H^2$, and so it can solved in closed form.  However, as the reduced
equation is first order, its solution depends on a single parameter,
and so in general only one of the two values $H(-\alpha)$ and
$H(-\beta)$ can be set equal to zero. The arguments above show that
the $z$-axis ODE for $V_1$ potentials in dMOND is a genuine Abel
equation of the second kind, to be solved numerically for assigned
density profile.

Concerning the asymptotic behavior of the density, following the same
treatment done for separable models in Sect. 3.1, it can be shown that
also regular $V_1$ models are characterized by a vanishing density at
the center. The asymptotic analysis at infinity reveals some
interesting difference with the cases in Sect. 3.2. In fact, let us
consider the $V_1$ family
\[
F(\tau)= {h(\tau)\ln\tau\over 2},
\quad h(\tau)\sim 1\quad {\rm for}\; \tau\to\infty,
\label{Fma1}
\]
similar to the family of finite mass systems discussed in Sect.~3.2.
We note that the parallel is only formal: while in the separable
models the potential at large radii becomes spherical, in this case it
remains ellipsoidal. This means that, at variance with the separable
case, $V_1$ models in the family above are characterized by infinite
mass (or, in case of finite mass, by space sectors with negative
density), because the potential of finite mass systems is dominated by
the monopole term at large radii.

We note that in Newtonian gravity eq.~(\ref{Fma1}) with $h=1$
corresponds to the Binney (1981) triaxial logarithmic potential,
$\ln(1+x^2/a^2 +y^2/b^2 +z^2/c^2)\propto\ln(\lambda\mu\nu)$, fully
discussed in ZP88 (Sects.~3.1 and 6.1 therein).  Here we found that
the Binney potential in dMOND leads to a non-spherical density
distribution of infinite mass, with a radial $r^{-3}$ profile at large
radii, with a negative density along the $z$-axis for sufficiently
large $\tau$, independently of the values of $a$, $b$, $c$. In
Cartesian coordinates:
\[
\rho(0,0,z)\sim {-2+c^2/a^2+b^2/a^2\over z^3},\quad 0<c\leq b\leq
a,\quad z\to\infty.
\label{eq:binneyz}
\]
What happens if we allow for a more general $h$ function in
eq.~(\ref{Fma1})? For simplicity we do not repeat the analysis done in
Sect.~3.2 for separable potentials, but we just report a few results.
First, we found that the negative densities at large radii along the
$z$-axis can be removed by appropriate choices of $A$ and $\epsilon$
in eq.~(\ref{eq:hlau}), but we were unable to construct everywhere
positive mass models in the family (\ref{Fma1}). Second, at variance
with the separable case, at large radii $\rho\propto 1/r^3$ (modulated
by an angular part) independently of $\epsilon$, so that {\it finite
  mass} dMOND systems in the $V_1$ family (\ref{Fma1}) do not exist.

Obviously, other choices of $F$ in eq.~(\ref{phi1}) are possible and
worth of investigation. For example, ZP88 (eq.~[2.13]) show that the
Newtonian potential of the density distribution
\[
\rho\propto {1\over 1+x^2/a^2+y^2/b^2+z^2/c^2},
\label{eq:rho1}
\]
belongs to the $V_1$, and that the gradient of the potential can be
written in terms of Carlson's (1979) symmetrized version of incomplete
elliptic integrals. These functions are a closed family under
differentiation, and therefore the dMOND density distribution
associated with this family can be written explicitely by using
functions no more complicated than elliptic integrals. Here, we do not
study these models.

\section{Summary and conclusions}

In this paper we studied the properties of density distributions
obtained from MOND potentials separable in ellipsoidal
coordinates. The investigation, besides the astrophysical context, is
also interesting because the MOND operator, in the weak field limit
(the so-called dMOND regime), reduces to the non-linear $p$-Laplace
operator $\nabla(\Vert\nabla\phi\Vert^{p-2}\nabla\phi)$ with $p=3$,
the critical case in $\Re^3$.

The main results obtained can be summarized as follows:

1. We proved that, as in the case of Newtonian gravity, also for MOND
systems with separable potentials in ellipsoidal coordinates, the
density profile along the $z$-axis (the long axis of the ellipsoidal
$\lambda$-surfaces) determines the potential and so the density
everywhere. The second-order ODE is however highly non-linear, and
probably unsolvable in closed form, even in dMOND, in the limit of
small flattenings and/or at large distances from the
origin. Therefore, while we formally proved that the Kuzmin property
holds in MOND, the associated Kuzmin formula is not known, and the
Kuzmin Theorem remains unproven.

2. However, we obtained rigorous asymptotic formulae for the density
near the origin and at infinity, in case of regular separable
potentials. We showed that, at variance with the case of Newtonian
separable systems, the density at the center vanishes, and we studied
the order of vanishing as a function of the order of regularity of the
potential. From this result it follows that MOND separable systems
with regular potentials are necessarily in the dMOND regime at their
center, so that they would appear as centrally dark matter dominated
(formally, with an infinite value for the mass-to-ligh ratio) when
interpreted in the context of Netwonian gravity. Of course, this
property may be shared by other sufficiently regular non-separable
potentials in MOND.

3. The analysis at large radii was performed under the assumptions of
finite total mass system and a separable potential sufficiently
regular to allow for a Frobenius expansion at infinity. In the regular
case (i.e., when the expansion of the shape function $h$ in
eq.~(\ref{eq:hlau}) reduces to a Taylor series in terms of integer
powers of $1/\tau$), the density at large radii is positive, provided
a certain coefficient in the expansion is greater than a negative
value (determined by the axial parameters of the ellipsoidal
coordinates). The density shape is not spherically symmetric, and the
radial decline is proportional to $\ln (r)/r^5$, at variance with the
Newtonian case of finite mass and finite flattening, when $\rho\propto
1/r^4$.  In case of a genuine Frobenius expansion with exponent
$0<\epsilon<1$, the total mass is still finite, but the density
profile at large radii is instead spherically symmetric, with a milder
radial decline $\rho\propto \ln (r)/r^{3+2\epsilon}$, and positive
provided the constant mentioned above is positive. This resembles the
Newtonian case of finite mass when the density profile along the
simmetry axis is steeper than $1/z^3$ but shallower than $1/z^4$.

4. We constructed some triaxial separable MOND models, everywhere
positive for all the explored axial ratios. The shapes range from
almost perfectly ellipsoidal systems to curious systems with density
depressions or overdensities (some of them similar to the models
constructed in Ciotti et al.~2006), depending on the specific choice
for the shape function $h$: these last models are unlikely to be
useful for the description of real stellar systems.

5. We briefly addressed the properties of the class of $V_1$
potentials introduced in ZP88: these potentials, albeit non-separable,
are simpler than the separable case, and they are known to obey the
Kuzmin theorem in Newtonian gravity. We showed that they obey the
Kuzmin property also in MOND, and we derived the ODE relating the
potential to the density profile along the $z$-axis. This equation is
considerably simpler than in the separable case, and in fact can be
transformed in a Abel equation of second kind. Unfortunately, the
general solution of this class of ODEs is not known, so the Kuzmin
theorem cannot be proved. As an example of $V_1$ systems, we studied
the case of the MOND analogue of the Binney (1981) logaritmic
potential, and we showed that the density profile becomes negative
along the $z$-axis, while the radial profile declines as $1/r^3$. Some
variants of this potential however admits positive densities (at large
radii) for some function $h$, but still declining as $1/r^3$, and so
being characterized by infinite total mass.

6. Finally, we showed (Appendix B) that power-law axisymmetric
potentials separable in parabolic coordinates, associated with a
central (weak) cusp can be constructed in MOND, in analogy with the
family discovered by Sridhar \& Touma (1997), albeit for a more
restricted range of central slopes. If a central black hole is added,
the central regions are still cuspy, but the cusp is Newtonian.

We conclude by noting a few points that are relevant for successive
investigations.  The first is related to the phase-space distribution
function. Presently our understanding of the phase-space distribution
function of MOND non-spherical systems still rely mostly on the
numerical Schwarzschild method (Wang et al. 2008, Wu et al. 2009,
2010) or N-body simulations (Nipoti et al. 2007ab, 2011), and the
resulting systems are expected to be non-separable. Here we can derive
some firm conclusion on MOND separable models. In fact, the vanishing
central density of triaxial regular models forces the associated
distribution functions to vanish for the values of the three integral
of motions allowing for orbits that cross the center.  Now, the orbit
classification for the Newtonian case assumes that the third
derivative of $F(\tau)$ is negative everywhere (e.g., Kuzmin 1973,
Hunter \& de Zeeuw 1992).  This is the case for the reference model
with $F(\tau)=-\tau^2\ln(\tau)/2$, leading to a third derivative which
is in fact $-1/\tau$ so indeed negative as we choose $\tau \geq
-\gamma \geq 0$. Moreover, in case of a Frobenius $h$ function, it is
easy to prove that the expansion coefficients can be chosen so that
negativity is assured (leaving true the positivity of the density). In
all these case, the models are supported by the four major orbit
families.  This indicates that these MOND models have the same orbit
structure as Newtonian separable systems. If the condition is
violated, then there are other/more orbit families
possible. Restricting to the first cases, it makes plausible that
selfconsistent models with vanishing central density might well exist
by populating the tube orbits only, leaving the box orbits out, as
these would be contributing positive density in the center.  The
machinery to construct such models with thin tubes only (zero radial
action) is available from Hunter \& de Zeeuw (1992): by leaving the
boxes out, this avoids having to compute the box orbit distribution
function by numerically solving a large set of linear equations. The
thin tube distribution functions are given in closed form, once
$F(\tau)$ is chosen and the density is known.

The second point is that the obtained results would change if the
$\mu$ function in eq.~(3) has a non-zero lower bound, i.e. $\mu(t)\to
\mu_0$ for $t\to 0$.  In fact, the deepest gravity regimes
observationally probed in isolated galaxies are $\approx 0.01\az$
(e.g., see Zhao 2007, Famaey et al. 2007, Wu et al. 2008). If such a
lower limit for $\mu$ exists, then the dynamics would be Newtonian at
very large radii and at very small radii, opening the possibility of
MOND St\"ackel models with finite total mass and non-zero central
density.

\section*{Acknowledgments}

We thank the Referee, Jin H. An,  for very useful comments.  H.Z. and
L.C. enjoyed the Oxford Problematik 2010 meeting, where aspects of
this work were discussed with S. Sridhar. L.C. was supported by the
MIUR grant PRIN2008; the warm hospitality of IPMU (Tokyo University)
and of Princeton University, where some part of this work was done, is
also aknowledged.

\appendix

\section{Ellipsoidal coordinates}

We report here the main properties of ellipsoidal coordinates relevant for the 
present work; for a full discussion and further references on the subject, 
see Z85 and ZLB85.

\subsection{Definitions}

Ellipsoidal coordinates $(\lambda, \mu, \nu)$  are curvilinear othogonal 
coordinates defined as the three real roots for $\tau$ of the cubic equation
\[
{x^2\over\tau +\alpha}+{y^2\over\tau +\beta}+{z^2\over\tau +\gamma}=1,
\label{eq:ellipse}
\]
where without loss of generality we assume $\alpha <\beta <\gamma<0$,
so that $0<-\gamma\leq\nu\leq -\beta\leq\mu\leq -\alpha\leq\lambda$.
Surfaces of constant $\lambda$ are ellipsoids, surfaces of constant
$\mu$ are hyperboloids of one sheet, and surfaces of constant $\nu$
are hyperboloids of two sheets\footnote{As some confusion often arise
  on this point, we remark that the $z$ axis is the {\it long} axis of
  the ellipsoidal $\lambda$-surfaces, but usually it is the {\it
    short} axis of triaxial mass models with St\"ackel potential
  (e.g., see Fig.~1 in ZLB85). See also Section 4 in de Zeeuw et
  al.~1986).}. For very large values of $\lambda$, the ellipsoidal
surfaces become more and more similar to spheres of radius
$\sqrt{\lambda}$.  The relations between $(\lambda,\mu,\nu)$ and the
Cartesian coordinates $(x,y,z)$ of a given point are
\begin{eqnarray}
\cases{
x^2=\displaystyle{{(\lambda +\alpha)(\mu +\alpha)(\nu +\alpha)\over 
     (\alpha -\beta)(\alpha -\gamma)},}\cr
y^2=\displaystyle{{(\lambda +\beta)(\mu +\beta)(\nu +\beta)\over 
     (\beta -\alpha)(\beta -\gamma)}},\cr
z^2=\displaystyle{{(\lambda +\gamma)(\mu +\gamma)(\nu +\gamma)\over 
     (\gamma -\alpha)(\gamma -\beta)}},
\label{eq:ellcoo}
}
\end{eqnarray}
so that the origin corresponds to $\lambda=-\alpha$, $\mu=-\beta$ and
$\nu=-\gamma$.
A series expansion shows that near the origin Cartesian and
ellipsoidal coordinates are related by the (first order) asymptotic
relations
\begin{eqnarray}
\cases{
x^2\sim\lambda+\alpha,\cr
y^2\sim\mu+\beta,\cr
z^2\sim\nu+\gamma.
\label{eq:centcor}
}
\end{eqnarray}
The metric coefficients are
\begin{eqnarray}
\cases{
\hl^2=\displaystyle{{(\lambda -\mu)(\lambda -\nu)\over 
     4 a_{\lambda}}},\cr
\hm^2=\displaystyle{{(\mu-\nu)(\mu -\lambda)\over 
     4 a_{\mu}}},\cr
\hn^2=\displaystyle{{(\nu -\lambda)(\nu -\mu)\over 
     4 a_{\nu}}},
}
\end{eqnarray}
where
\[
a_{\tau}=(\tau +\alpha)(\tau +\beta)(\tau +\gamma);
\]
note that $a_{\lambda}>0$, $a_{\nu}>0$, but $a_{\mu}<0$, consistent
with the positivity of the metric coefficients.  The gradient operator
reads
\[
\nabla={\evl\over\hl}{\partial\over\partiall}+
       {\evm\over\hm}{\partial\over\partialm}+
       {\evn\over\hn}{\partial\over\partialn},
\label{gradel}
\]
where $(\evl,\evm,\evn)$ is the local basis of mutually orthogonal
unitary vectors (e.g., Arfken \& Weber 2005).  Therefore, the squared
norm of the gravitational field becomes
\[
||\nabla\phi ||^2={1\over\hl^2}\left({\partial\phi\over\partiall}\right)^2+
                  {1\over\hm^2}\left({\partial\phi\over\partialm}\right)^2+
                  {1\over\hn^2}\left({\partial\phi\over\partialn}\right)^2,
\label{gradels}
\]
while the differential operator $\Dphi$ in eq.~(\ref{dMONDell}) is 
\[
\Dphi={1\over\hl^2}{\partial\phi\over\partiall}{\partial\over\partiall}+
         {1\over\hm^2}{\partial\phi\over\partialm}{\partial\over\partialm}+
         {1\over\hn^2}{\partial\phi\over\partialn}{\partial\over\partialn}.
\label{eq:dphi}
\]
Finally, the Laplace operator can be written as
\[
\nabla^2 =\nablasl+\nablasm+\nablasn,
\label{nablas}
\]
where
\[
\nablasl = {2\over (\lambda -\mu)(\lambda -\nu)}\left[
2 a_{\lambda}{\partial^2\over\partiall^2}+
{\partial a_{\lambda}\over\partiall}{\partial\over\partiall}
\right],
\label{nablasl}
\]
and $\nablasm$ and $\nablasn$ follow from the equation above by the
rotation $\lambda\to\mu\to\nu\to\lambda$, applied once and twice,
respectively.

\subsection{The leading terms of density expansion at the center}

With heavy but straightforward computation it can be shown that the
coefficients $\Ao$, $\Bo$, and $\Co$ appearing in
eqs.~(\ref{eq:normcen1})-(\ref{eq:dnormcen1}), needed in the
density expansion near the center of generic separable dMOND system,
are
\[
\Ao={\beta-\gamma\over \triangle^2}A_1,\quad
\Bo={\alpha-\gamma\over\triangle^2}B_1,\quad
\Co={\alpha-\beta \over\triangle^2}C_1,
\label{eq:coecen1a}
\]
where $\triangle$ is defined in eq.~(\ref{eq:tria}), and
\begin{eqnarray}
\cases{
A_1=F(-\alpha)(\beta+\gamma-2\alpha)(\beta-\gamma)-F'(-\alpha)\triangle+\cr
\quad\quad F(-\beta)(\alpha-\gamma)^2-F(-\gamma)(\alpha-\beta)^2,\cr
B_1=F(-\beta)(\alpha+\gamma-2\beta)(\alpha-\gamma)+F'(-\beta)\triangle+\cr
\quad\quad F(-\alpha)(\beta-\gamma)^2-F(-\gamma)(\alpha-\beta)^2,\cr
C_1=F(-\gamma)(\alpha+\beta-2\gamma)(\alpha-\beta)-F'(-\gamma)\triangle+\cr
\quad\quad F(-\alpha)(\beta-\gamma)^2-F(-\beta)(\alpha-\gamma)^2.
\label{eq:coecen1b}
}
\end{eqnarray}

In the special case $\Ao=\Bo=\Co=0$, the values of $F'$ at the center
are fixed by the vanishing of the system above. The coefficients of
the resulting expansions reported in eq.~(\ref{eq:dencen2}) are 
\[
\Ad=-{\beta-\gamma\over \triangle^2}A_2,\quad
\Bd={\alpha-\gamma\over\triangle^2}B_2,\quad
\Cd=-{\alpha-\beta \over\triangle^2}C_2,
\label{eq:coecen2a}
\]
where now
\begin{eqnarray}
\cases{
A_2=F''(-\alpha)\triangle -2F(-\alpha)(\beta-\gamma)+\cr
\quad\quad 2F(-\beta)(\alpha-\gamma)-2F(-\gamma)(\alpha-\beta),\cr
B_2=F''(-\beta)\triangle -2F(-\alpha)(\beta-\gamma)+\cr
\quad\quad 2F(-\beta)(\alpha-\gamma)-2F(-\gamma)(\alpha-\beta),\cr
C_2=F''(-\gamma)\triangle -2F(-\alpha)(\beta-\gamma)+\cr
\quad\quad 2F(-\beta)(\alpha-\gamma)-2F(-\gamma)(\alpha-\beta).
}
\label{eq:coecen2b}
\end{eqnarray}

The additional request of regularity, i.e., $\Ad=\Bd=\Cd=0$, fixes the
values of $F''$ at the center from the vanishing of the system above. The coefficients of
the resulting expansions reported in eq.~(\ref{eq:dencen3}) are 
\begin{eqnarray}
\cases{
\displaystyle{\At=-{ \beta-\gamma\over \triangle} F'''(-\alpha)   },\cr
\displaystyle{\Bt={\alpha-\gamma\over\triangle} F'''(-\beta)  },\cr
\displaystyle{\Ct=-{\alpha-\beta \over\triangle}  F'''(-\gamma) }.
}
\label{eq:coecen3}
\end{eqnarray}

\subsection{Cartesian planes in ellipsoidal coordinates}

In the study of the shape and density distribution of triaxial mass
models expressed in ellipsoidal coordinates it may be helpful to have
the expression for the coordinate planes $(x,y,0)$, $(x,0,z)$, and
$(0,y,z)$ in terms of the ellipsoidal coordinates. The following
formulae can be easily deduced by simple geometrical arguments (see
also de Zeeuw et al. 1986). 

The $(x,y)$ plane is obtained by requiring that $z=0$ in
eq.~(\ref{eq:ellcoo}). This request leads to fix $\nu=-\gamma$, and
solve for assigned $x^2$ and $y^2$ the resulting system for
$L\equiv\lambda+\alpha\geq 0$ and $M\equiv\mu+\beta\geq 0$. Therefore,
\[
\rho(x,y,0)=\rho(L-\alpha,M-\beta,-\gamma),
\]
where the density at the l.h.s. is intended to be expressed in Cartesian coordinates,  
\[
L={R^2-\delta+\sqrt{(R^2-\delta)^2+4\delta x^2}\over 2},\quad
\delta\geq\beta-\alpha>0,
\label{eqAL}
\]
and $M=R^2-L$, $R^2=x^2+y^2$.

The situation is slightly more complicated for the other two planes. In
fact, the $(x,z)$ plane is fully covered by the union of the region
\[
S_{\mu}\equiv \left\{
(x,z):\;\; {z^2\over\gamma-\beta}-{x^2\over\beta-\alpha}\leq 1
\right\},
\]
obtained by fixing $\mu=-\beta$ in eq.~(\ref{eq:ellcoo}), with the
complementary region $S_{\nu}\equiv \Re^2/S_{\mu}$, obtained for
$\nu=-\beta$.  Solving the resulting systems and asking for positivity
of the functions $L\equiv\lambda+\alpha$ and $N\equiv\nu+\gamma$ (in
$S_{\mu}$), or $M\equiv\mu+\gamma$ (in $S_{\nu}$), one gets
\begin{eqnarray}
\rho(x,0,z)=\cases{
\rho(L-\alpha,-\beta,N-\gamma),\quad (x,z) \in S_{\mu},\cr
\rho(L-\alpha,M-\gamma,-\beta),\quad (x,z) \in S_{\nu},
}
\end{eqnarray}
where $L$ is again given by eq.~(\ref{eqAL}) but now $R^2=x^2+z^2$, 
$\delta=\gamma-\alpha\geq 0$, and $M=N=R^2-L$.

Finally, a similar analysis shows that the $(y,z)$ plane is also
separated in two regions,
\[
S_{\lambda}\equiv \left\{
(y,z):\;\; {y^2\over\beta -\alpha}+{z^2\over\gamma-\alpha}\leq 1
\right\},
\]
obtained by fixing $\lambda=-\alpha$ in eq.~(\ref{eq:ellcoo}), and the
complementary region $S_{\mu}\equiv \Re^2/S_{\lambda}$, obtained for
$\mu=-\alpha$.  Solving the resulting systems and asking for
positivity of the functions $N\equiv\nu+\gamma$ and $M\equiv\mu+\beta$
(in $S_{\lambda}$), or $L\equiv\lambda+\beta$ (in $S_{\mu}$), we now
get
\begin{eqnarray}
\rho(0,y,z)=\cases{
\rho(-\alpha,M-\beta,N-\gamma),\quad (y,z) \in S_{\lambda},\cr
\rho(L-\beta,-\alpha,N-\gamma),\quad (y,z) \in S_{\mu},
}
\end{eqnarray}
where
\[
L=M={R^2-\delta+\sqrt{(R^2-\delta)^2+4\delta y^2}\over 2},\quad
\delta=\gamma-\beta\geq 0,
\]
$R^2=y^2+z^2$, and $N=R^2-M$. 

\begin{figure*}
\centerline{
\psfig{file=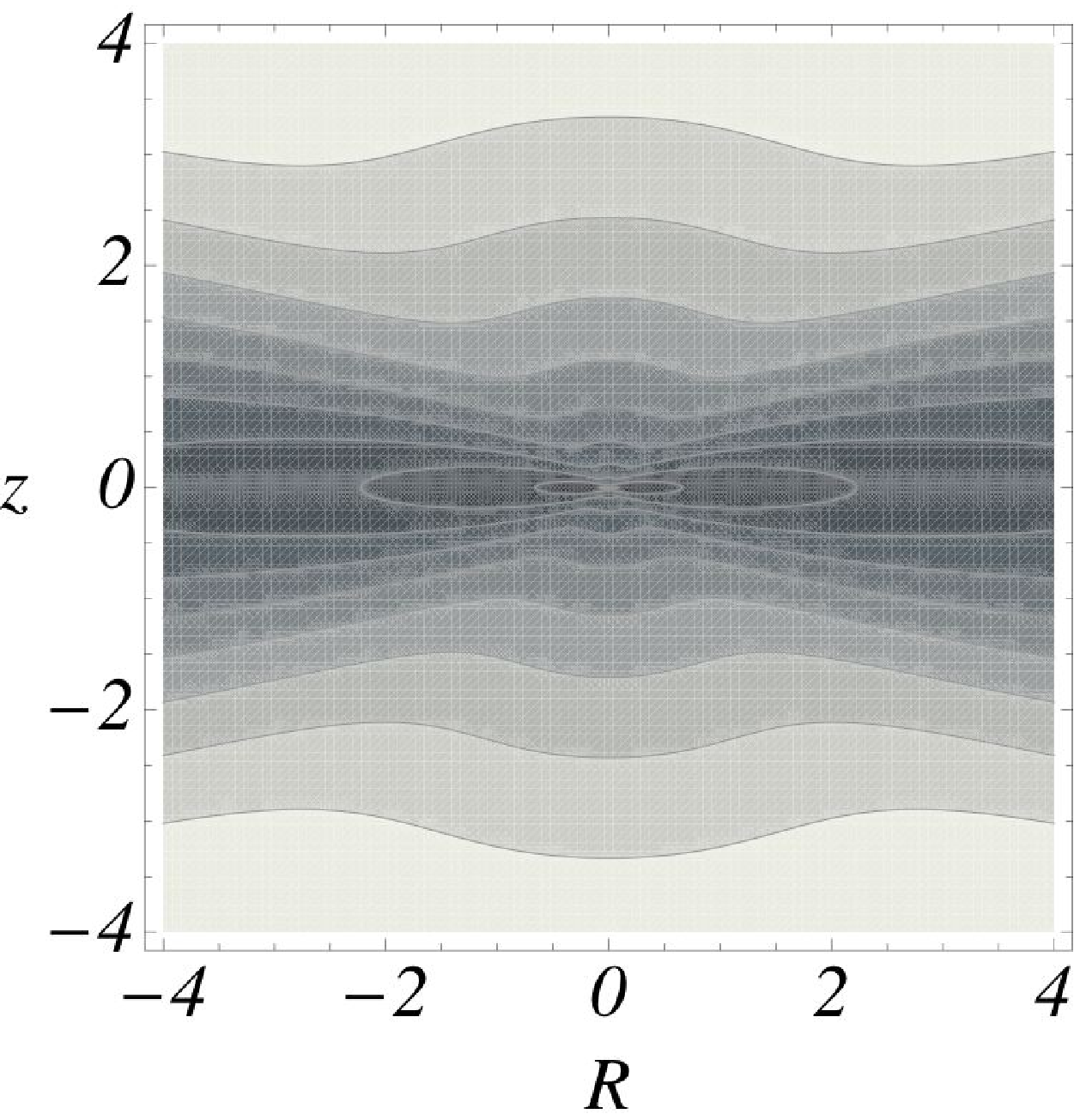,width=0.31\hsize}
\psfig{file=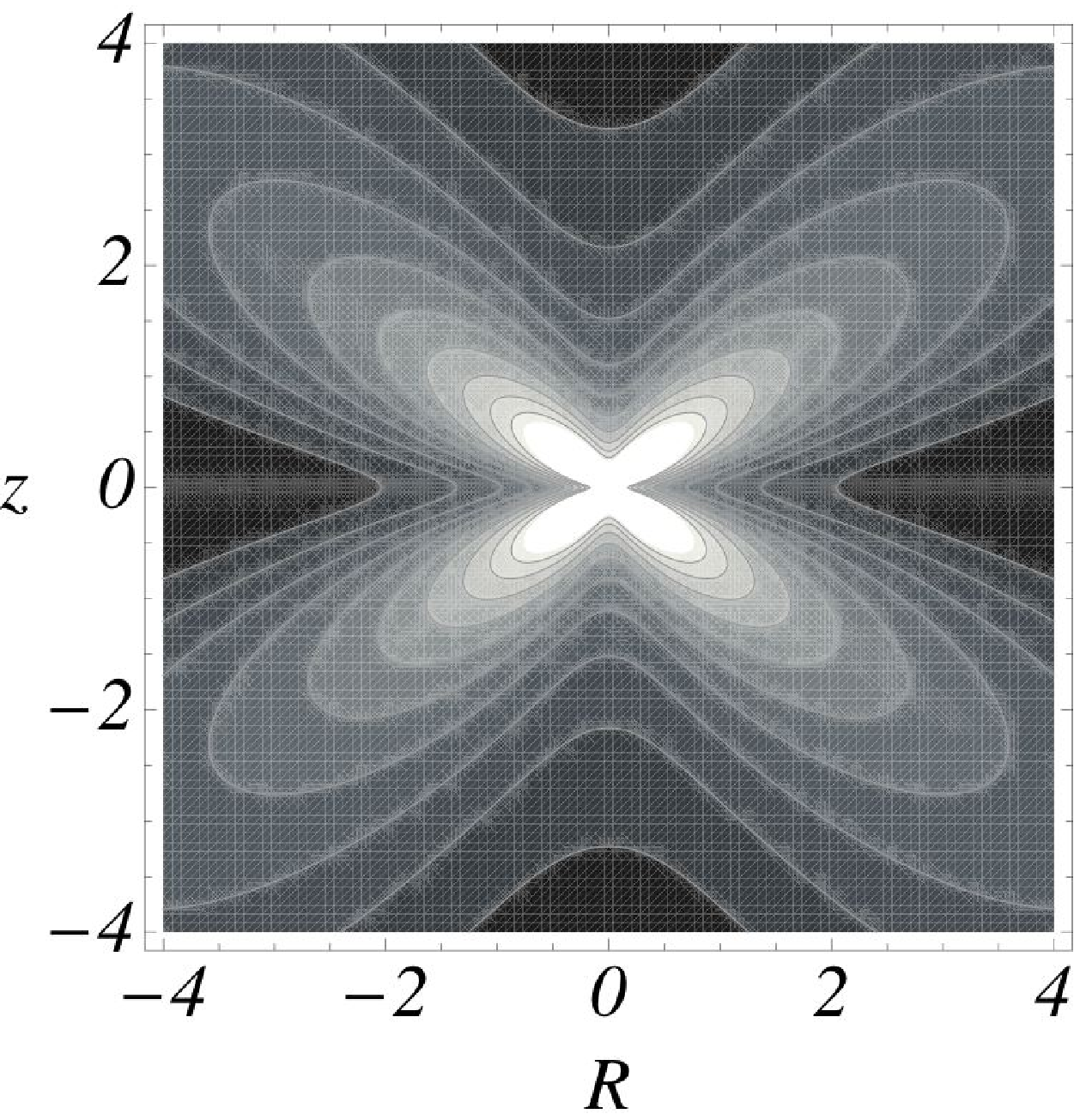,width=0.31\hsize}
\psfig{file=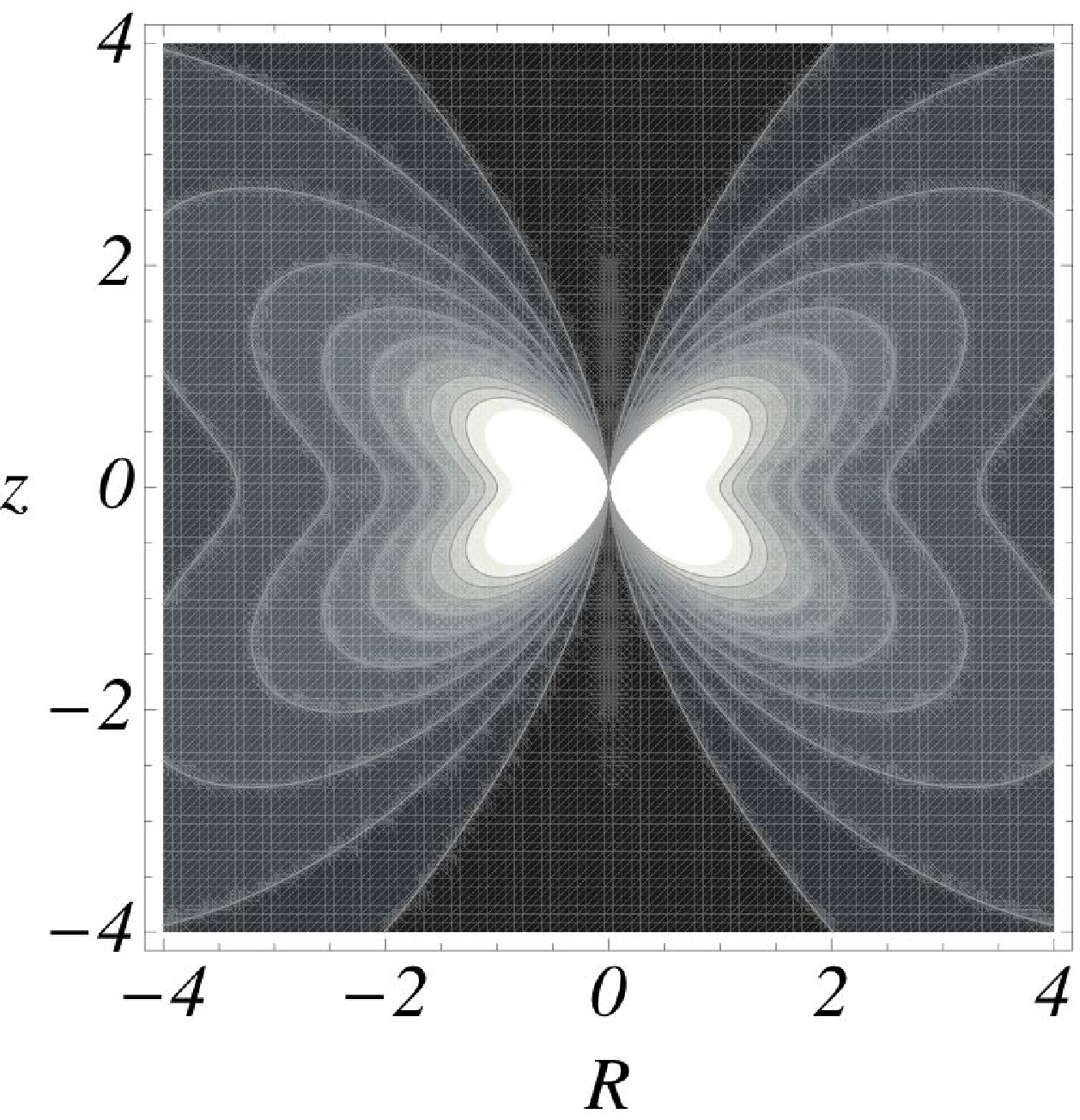,width=0.31\hsize}
}
\caption{Isodensity contours of the dMOND analogues of the Sridhar \&
Touma (1997) power-law axisymmetric cuspy models with separable
potential in parabolic coordinates. From left to right, $k=1/3$,
$2/3$, and $1$. As explained in the test, the density diverges at the
origin for $1/2<k\leq 1$, while it vanishes for $0<k<1/2$. For
$k=1/2$ the density (not shown) is independent of $r$, i.e., it is
constant on cones with the common vertex on the origin.}
\end{figure*}

\section{The power-law axisymmetric separable model}

A property common to all triaxial systems with a Newtonian potential
separable in ellipsoidal coordinates is a constant density core.  In
the axisymmetric case, Sridhar \& Touma (1997) were able to construct
separable potentials supporting a central density cusp.  The question
is whether such property carries on in MOND as well. We show that this
is in fact the case.  We start from the separable potential
\begin{eqnarray}
\phi(r_{+},r_{-}) &=& 2{r_{+}^{3-k} -  |r_{-}|^{3-k}\over
r_{+} -  r_{-}}=\cr
&&
r^{2-k}\left[(1+\cos\theta)^{3-k} + (1-\cos\theta)^{3-k}\right],
\end{eqnarray}
where the parabolic coordinates $r_+$ and $r_-$ are related to the
standard spherical coordinates by the identities
\[
r_{+}   = r (1+\cos\theta), \quad r_{-}= r (1-\cos\theta).
\]
Sridhar \& Touma (1997) show that the density distribution associated
with the potential above, via the Laplace operator, is
\[
\rho\propto {(2-k-\cos\theta)(1+\cos\theta)^{2-k}+
(2-k+\cos\theta)(1-\cos\theta)^{2-k}\over r^k},
\]
and discuss its properties as a function of $k$. In
particular, for $0<k<1$ the density is cusped at the origin and
positive everywhere, and so they conclude that cuspy systems (of infinite mass) with
separable potential exist, at least in the axisymmetric case. In the
critical case, $k=1$, the density is cuspy, but it vanishes on the
$z$-axis, being $\rho\propto \sin^2\theta/r$.

We do not embark on the interesting but long discussion of the density
related to the potential (B1) in MOND, but we note the following
results. First, it is easy to show that the force depends on radius as
$r^{1-k}$, and does not vanish along any direction (as the radial
component of $\nabla\phi$ never vanishes).  It follows that for $k>1$
the MOND system will be similar - near the origin - to the Newtonian
case, when however the density is unphysical (Sridhar \& Touma
1997). For $0<k<1$ the system is in the dMOND regime near the origin
(and Newtonian at infinity), while in the critical $k=1$ case the
force is independent of $r$, and so the system can be constructed in the 
dMOND regime everywhere. An application of the $3$-Laplace operator to
the family of power-law potential with $0<k\leq 1$ easily shows that
the density near the center (or everywhere, in the $k=1$ case), can be
written as
\[
\rho\propto r^{1-2k}f_k(\theta),
\]
where the explicit function $f_k$ is easily calculated but is not
reported here. Remarkably, in the range $0<k\leq 1$ the function $f_k$
is nowhere negative. From eq.~(B4) it follows that the density
vanishes at the center for $0<k<1/2$, is independent of $r$ for
$k=1/2$ (but with different values along different directions), and it
is cusped for $1/2 <k\leq 1$. Moreover, $f_1\propto\sin^2\theta$, as
in the Newtonian case. Therefore, separable potentials with a central
(weak) cusp can be constructed also in MOND, but only in the
restricted range $1/2<k<1$. However, the resulting densities, albeit
positive, are still characterized by an infinite total mass, and their
shapes are quite unnatural, as can be seen from Fig.~A1, where some
examples are presented. Sridhar \& Touma (1997) also showed that a
black hole can be added at the center of these models, leaving
separability unaffected. This remains true in MOND, as the
gravitational field of the black hole switches the field near the
center from dMOND to Newtonian, so that the system will be cuspy also
for $0<k\leq 1/2$: however, this is not a ``genuine'' MOND cusp.

We finally note that the superposition of potentials of the family
(B1) with different values of $k$ is still a separable potential. Of
course, the dMOND operator is non-linear, so that the associated
densities are not the sum of the separate components. However, we
performed some numerical experiments, and we found that the resulting
densities (for $0<k\leq 1$) are still positive. This could open the
way to the construction of new families of axisymmetric MOND systems
with separable potentials and more general density distributions than
pure power-laws.

\end{document}